\documentclass[twocolumn,resetfootnote]{aastex7}

\usepackage{graphicx} % Required for inserting images
\usepackage{amsmath}
\usepackage{soul}

\newcommand{\be}{\begin{equation}}
\newcommand{\ee}{\end{equation}}
\newcommand{\bel}[1]{\begin{equation}\label{#1}}
\newcommand{\ba}{\begin{eqnarray}}
\newcommand{\ea}{\end{eqnarray}}
\newcommand{\bal}[1]{\begin{eqnarray}\label{#1}}

\newcommand{\myhdashline}{%
  \noalign{%
    \hbox to \textwidth{%
      \rule{3pt}{\arrayrulewidth}%
      \xleaders\hbox{\hskip3pt\rule{3pt}{\arrayrulewidth}}\hfill
      \kern0pt
    }%
  }%
}

\begin{document}

\title{On Bimodality in the Eccentricity Distribution of Galactic Double Neutron Stars}

\author[0000-0002-2215-1841]{Aldana Grichener}
\affiliation{Steward Observatory and Department of Astronomy, University of Arizona, 933 North Cherry Avenue, Tucson, AZ 85721, USA}
\email{agrichener@arizona.edu}
\correspondingauthor{Aldana Grichener}
\email{agrichener@arizona.edu}

\author[0000-0002-0492-4089]{Paul Disberg}
\affiliation{School of Physics and Astronomy, Monash University, Clayton, Victoria 3800, Australia}
\affiliation{The ARC Centre of Excellence for Gravitational Wave Discovery—OzGrav, Australia}
\email{ paul.disberg@monash.edu}

\author[0000-0002-8032-8174]{Ryosuke Hirai}
\affiliation{Astrophysical Big Bang Laboratory (ABBL), RIKEN Pioneering Research Institute (PRI), 2–1 Hirosawa, Wako, Saitama 351-0198, Japan}
\affiliation{School of Physics and Astronomy, Monash University, Clayton, Victoria 3800, Australia}
\affiliation{The ARC Centre of Excellence for Gravitational Wave Discovery—OzGrav, Australia}
\email{ryosuke.hirai@riken.jp}

\author[0000-0002-6134-8946]{Ilya Mandel}
\affiliation{School of Physics and Astronomy, Monash University, Clayton, Victoria 3800, Australia}
\affiliation{The ARC Centre of Excellence for Gravitational Wave Discovery—OzGrav, Australia}
\email{ilya.mandel@monash.edu}

\begin{abstract}
The detection of Galactic double neutron stars (DNSs) through pulsar timing offers a unique opportunity to probe massive stellar and binary evolution. The current observed DNS population shows a scarcity of systems at intermediate eccentricities, although the present sample does not provide statistically significant evidence for an underlying bimodal distribution. We explore whether such a feature can arise naturally from DNS formation physics and serve as a prediction for future observations. We propose that this possible bimodality can result from a non-monotonic relationship between progenitor mass and neutron star (NS) mass, provided that the second-born NS receives a sufficiently small natal kick. We illustrate this scenario using the population synthesis code \textsc{compas}, and find that our DNS population model produces an intermediate-eccentricity deficit broadly similar to that suggested by current observations and can reproduce the observed orbital period--eccentricity distribution relatively well. We also show that most measured second-born NS masses fall within the ranges predicted by our model at the corresponding DNS orbital eccentricities, providing a complementary observational diagnostic that is less sensitive to uncertainties in the preceding binary evolution.
Our results therefore predict that an eccentricity bimodality should become more apparent as the Galactic DNS sample grows. 
\end{abstract}

% ==========================================================
\section{Introduction}
\label{sec:Intro}
% ==========================================================

Double neutron star (DNS) systems are central to massive-star research, and can provide insights into stellar evolution, supernova (SN) explosions, and related transients. Galactic DNSs at orbital periods ranging from just under 2 hours to 45 days have been observed through radio pulsar timing. Since the discovery of the Hulse-Taylor system five decades ago \citep{HulseTaylor:1975}, the sample of observed galactic DNSs has been significantly expanding, and is currently composed of $\rm 29$ binaries (see Table~\ref{tab1}), with an average of one new detection per year over the past decade. More than $40\%$ of these systems are expected to merge within the Hubble time \citep{Bernadich:2023}. This sample provides the best observational constraint to date on the origin of DNSs.       

DNSs are expected to form through isolated evolution of massive binary systems (e.g., \citealt{FlanneryvdH:1975}; \citealt{Tauris:2017}; \citealt{VignaGomez:2018}; \citealt{Grichener:2023}). They typically begin their evolution from two massive main sequence (MS) stars orbiting their common center of mass. The more massive star evolves into a red supergiant (RSG) and interacts with its companion via mass transfer. It later experiences a SN explosion, leaving a neutron star (NS) behind. The newly born NS might receive a natal kick due to asymmetries in the explosion (e.g., \citealt{LyneLorimer:1994}) and a Blaauw kick caused by instantaneous mass loss during core collapse \citep{Blaauw:1961}, which could potentially disrupt the binary system. If the binary remains bound after the explosion, a common envelope evolution (CEE) stage \citep{Ivanova:2020} can occur when the initially lighter companion evolves into a RSG, driving the NS and its core closer together due to dynamical friction. Unless they merge during this CEE phase (e.g., \citealt{Grichener:2025}), the surviving core can explode in a second SN event. If the system is not disrupted by the explosion, a DNS binary is born. Therefore, the assumptions regarding the masses and kicks of the NSs formed in the SN explosion, and the treatment of binary mass transfer and CEE, are crucial for studying DNS binaries, both at the single system and population levels.

The formation and evolution of DNSs have been explored in dozens of population synthesis studies (see \citealt{MandelBroekgaarden:2021} for a review), whose predictions for the merger rates of DNSs span several orders of magnitude and are often compared with gravitational-wave (GW) observations. The growing sample of Galactic DNSs discovered through radio pulsar observations provides complementary constraints on DNS formation through their observed orbital properties \citep[e.g.,][]{AndrewsMandel:2019,Mandel:2020,Chattaraj:2025,Nair:2025,Disberg:2026b}. In the current Galactic sample 21 systems exhibit low measured eccentricities of $e_{\rm m} \lesssim 0.4$, while the remaining eight are characterized by high eccentricities of $e_{\rm m} \gtrsim 0.58$, leaving the intermediate range in the present-day sample unpopulated, and hinting at a possible bimodality in the birth eccentricity distribution, which has been a topic of discussion \citep[e.g.,][]{vanDenHeuvel:2007,BeniaminiPiran:2016,AndrewsMandel:2019}.  Given the small number of observed systems, however, it remains unclear whether this scarcity reflects an underlying bimodal distribution or simply arises from small number statistics. Despite this uncertainty, the possibility of such a bimodality is motivated by previous indications of bimodal or multi-component structure associated with the formation of NSs and their binary systems, such as the orbital properties of high mass X-ray binaries \citep[e.g.,][]{Pfahl:2002}  and Be X-ray binaries \citep{Knigge:2011,Valli:2025}, and the velocities of young pulsars \citep{Igoshev:2020}. 

In this study, we argue that an eccentricity bimodality in DNS eccentricities could arise naturally from a non-monotonic relationship between progenitor mass and NS mass, combined with weak, potentially negligible natal kicks associated with the ultra-stripped supernovae (USSNe) that form the second NS in close binaries. The break in the remnant mass function (see left panel of Fig.~\ref{fig:two_panels_COcore_vs_MNS2_and_beta_vs_e}, for instance) would translate into a non-monotonic distribution of mass lost during the second SN, and hence into a potentially bimodal distribution of Blaauw kicks. If the natal kicks are small, this mechanism can produce an eccentricity bimodality similar to that motivated by the current Galactic DNS observations, while making a testable prediction for future samples \citep[see also][]{Disberg:2026b}. We present the observed sample in Section~\ref{sec:observations}, motivate our model and illustrate it with population synthesis in Section~\ref{sec:BimodalEccentricity}, discuss observational constraints on the second SN kick in Section~\ref{sec:SecondNSkick}, address potential caveats and alternative solutions in Section~\ref{sec:discussion}, and summarize in Section~\ref{sec:summary}.

% ==========================================================
\section{DNS observations}
\label{sec:observations}
% ==========================================================

As our sample of observed Galactic DNS systems, we consider the pulsars listed in the Australia Telescope National Facility (ATNF) Pulsar Catalog\footnote{\href{https://www.atnf.csiro.au/research/pulsar/psrcat/}{https://www.atnf.csiro.au/research/pulsar/psrcat/}.} \citep{Manchester:2005}, and select those with companions that are labeled as NSs, even if their nature is uncertain. We classify systems as `Observed DNSs' when the companion is generally accepted to be a NS and the system does not have a correlation with a globular cluster (GC), and as `potential/GC DNSs' if there is discussion in the literature about the nature of the pulsar's companion \citep[e.g.][]{Keith:2009,Van_Leeuwen:2015,Ng:2018b,Barr:2024}, or if the DNS is associated with a GC. In the latter case, the NSs or their progenitors may have been influenced by dynamical interactions \citep[e.g.][]{Phinney:1991}, and therefore cannot be directly compared to models of isolated binary evolution. It is possible that some field DNSs without a GC association, particularly high-eccentricity ones, were originally formed in star clusters and were later ejected \citep{AndrewsMandel:2019} (though \citealt{Ye:2019} argue that the rate of this channel is likely too low). We include the `potential/GC DNSs' in our qualitative analysis, since they follow similar trends in orbital period and eccentricity \citep[see also, e.g.,][]{Valli:2025}, but note their different or uncertain origin in our quantitative estimates. 

From the ATNF catalog, we obtain the observed orbital period, eccentricity, and characteristic spin-down age. Since the binary orbits have changed due to the emission of GWs, we use the model of \citet{Peters:1964} to integrate these orbits back in time for a duration equal to the characteristic age of the observed radio pulsar and estimate the orbital period and eccentricity immediately after the second SN. For this integration, we take NS masses of $1.4M_{\odot}$, but note that using mass estimates from the literature \citep[e.g.,][]{Ding:2024} would not noticeably change our results. This method assumes that the characteristic age of the observed pulsar approximately equals the time since the formation of the DNS. Although (1) the characteristic age does not always accurately reflect the true pulsar age \citep[e.g.][]{lyne:1996}, and (2) the observed pulsar is typically the recycled primary and thus slightly older than the age of the DNS, characteristic age estimates are generally consistent with kinematic ages \citep{Igoshev:2019,Disberg:2024b,Disberg:2025}, and are therefore usually considered reasonable pulsar age proxies (e.g., \citealt{Maoz:2024}). 

We adopt the total mass measurements from the literature (see Table~\ref{tab1} for references), and for systems with well-constrained component masses, we identify the recycled NS as the first-born object, since it was spun up through mass transfer from the progenitor of the second-formed NS. J1906+0746 is an exception, as the observed pulsar is young and unrecycled, and is therefore cataloged as the second-born NS. We include the component mass measurements of DNSs associated with GCs, but note that past dynamical interactions can make their evolutionary ordering uncertain. 

In Table~\ref{tab1}, we list the measured orbital periods and eccentricities together with their back-integrated values, characteristic ages, and mass estimates. While the measurement uncertainties are relatively small, with $\log_{10}(\sigma_{P_{\rm m}}/{\rm days})\lesssim -6$ for the periods and $\log_{10}(\sigma_{e_{\rm m}})\lesssim -4$ for the eccentricities, the back-integrated values are considerably more uncertain, especially for short-period systems. The uncertainties in mass measurements vary more substantially (e.g., \citealt{Ding:2024}).

\begin{table*}
\centering
\scriptsize
\caption{Properties of Galactic DNSs. For each system we list the measured orbital period ($P_{\rm m}$) and eccentricity ($e_{\rm m}$), followed by the period ($P_{\rm b}$) and eccentricity ($e_{\rm b}$) {\it immediately after the second SN}. We also present their characteristic age ($\tau_{c}$), total mass estimate ($M_{\rm T}$), mass of the first-born NS ($M_{\rm NS,first}$) and mass of the second-born NS ($M_{\rm NS,second}$) when well constrained, as well as the references for the mass estimates. The upper block shows the `observed DNSs' whose classification is generally agreed upon in the literature, and the lower block lists the `potential/GC DNSs',with the GC DNSs above the dashed line and the potential DNSs below it.} \label{tab1}
\begin{tabular*}{\textwidth}{@{\extracolsep{\fill}}lccccccccc@{}}
Name & $P_{\rm m}$ & $e_{\rm m}$ & $P_{\rm b}$ & $e_{\rm b}$ & $\tau_{c}$ & $M_{\rm T}$ & $M_{\rm NS,first}$ & $M_{\rm NS,second}$ &  ref \\
 & [days] & & [days] & & [Myr] & [$M_{\odot}$] & [$M_{\odot}$] & [$M_{\odot}$] & \\\hline
J0453+1559 & 4.07 & 0.113 & 4.08 & 0.113 & 3900 & 2.73 & 1.56 & 1.17 & (1)\\
J0509+3801 & 0.380 & 0.586 & 0.459 & 0.627 & 153 & 2.81 & 1.40 & 1.41 & (2)\\
J0641+0448 & 3.73 & 0.145 & 3.74 & 0.146 & 6200 & 2.59 & 1.32 & 1.27 & (3)\\
J0737--3039A & 0.102 & 0.088 & 0.171 & 0.147 & 204 & 2.59 & 1.34 & 1.25 & (1)\\
J1155--6529 & 3.69 & 0.260 & 3.68 & 0.260 & 3574 & - & - & - & \\
J1208--5936 & 0.632 & 0.348 & 1.09 & 0.483 & ${>}$10000$^{\text{*}}$ & 2.59 & - & - & (1)\\
J1325--6253 & 1.82 & 0.064 & 1.86 & 0.065 & 9560 & 2.57 & - & - & (1)\\
J1411+2551 & 2.62 & 0.170 & 2.64 & 0.172 & 10300 & 2.54 & - & - & (1)\\
J1518+4904 & 8.63 & 0.249 & 8.64 & 0.250 & 23900$^{\dagger}$ & 2.72 & 1.47 & 1.25 & (4)\\
B1534+12 & 0.421 & 0.274 & 0.439 & 0.283 & 248 & 2.68 & 1.33 & 1.35 & (1)\\
J1756--2251 & 0.320 & 0.181 & 0.357 & 0.200 & 443 & 2.57 & 1.34 & 1.23 & (1)\\
J1757--1854 & 0.183 & 0.606 & 0.490 & 0.777 & 130 & 2.73 & 1.34 & 1.39 & (1)\\
J1811--1736 & 18.78 & 0.828 & 18.80 & 0.828 & 1830 & 2.57 & - & - & (1)\\
J1829+2456 & 1.18 & 0.139 & 1.29 & 0.152 & 12400 & 2.61 & 1.31 & 1.30 & (1)\\
J1846--0513 & 0.613 & 0.209 & 0.625 & 0.213 & 363 & 2.63 & - & - & (1,5)\\
J1901+0658 & 14.5 & 0.366 & 14.5 & 0.366 & 5500 & 2.79 & - & - & (1)\\ 
J1913+1102 & 0.206 & 0.090 & 0.432 & 0.186 & 2690 & 2.89 & 1.60 & 1.29 & (6)\\
B1913+16 & 0.323 & 0.617 & 0.420 &  0.669 & 109 & 2.83 & 1.44 & 1.39 & (1)\\
J1930--1852 & 45.1 & 0.399 & 45.1 & 0.399 & 163 & 2.59 & - & - & (1)\\
J1946+2052 & 0.078 & 0.064 & 0.182 & 0.150 & 292 & 2.53 & 1.28 & 1.25 & (7)\\\hline
J0514--4002E & 7.45 & 0.708 & 7.45 & 0.708 & 459 & 3.89 & - & - & (8)\\
J1807--2500B & 9.96 & 0.747 & 9.97 & 0.747 & 806 & 2.57 & 1.36 & 1.21 & (9) \\
B2127+11C & 0.335 & 0.681 & 0.493 & 0.746 & 97 & 2.71 & 1.35 & 1.36 & (10)\\ 
\myhdashline
J1018--1523 & 8.98 & 0.228 & 8.99 & 0.228 & 12100 & 2.30 & - & - & (1)\\
J1753--2240 & 13.6 & 0.304 & 13.6 & 0.304 & 1550 & - & - & - & \\
J1755--2550 & 9.70 & 0.089 & 9.70 & 0.089 & 2.05 & - & - & - & \\
J1759+5036 & 2.04 & 0.308 & 2.11 & 0.316 & 11500 & 2.68 & - & - & (2)\\
J1906+0746 & 0.166 & 0.085 & 0.166 & 0.085 & 0.113 &  2.61 & 1.30 & 1.31 & (11)\\
J2150+3427 & 10.6 & 0.601 & 10.6 & 0.602 & 2880 & 2.59 & - & - & (1)\\\hline
\end{tabular*}
\tablecomments{\footnotesize The measured periods ($P_{\rm m}$), eccentricities ($e_{\rm m}$), and ages ($\tau_{\rm c}$) were taken from the ATNF Pulsar Catalog \citet{Manchester:2005}, and the listed values of $P_{\rm b}$ and $e_{\rm b}$ are the result of integrating the orbits back in time following the formalism of \citet{Peters:1964} to correct for the effects of GW emission. We assume NS masses of $1.4M_{\odot}$ for the integration, but note that this has negligible effect on the resulting values compared to adopting the total mass estimates. (*) We use an age of 10 Gyr to determine the birth properties, but changing this to 14 Gyr does not affect the results significantly. ($\dagger$) Age exceeds a Hubble time, we therefore use an age of $14$ Gyr for the back integration. We note that this does not affect the period and eccentricity estimates. References for mass measurements: (1) Catalog of \citet{Chattaraj:2025} and references therein; (2) \citet{McEwen:2024}; (3) \citet{Yang:2026}; (4) \citet{Tan:2024}; (5) \citet{Zhao:2024}; (6) \citet{Miao:2026}; (7) \citet{Meng:2025}; (8) \citet{Barr:2024}; (9) \citet{Lynch:2012}; (10) \citet{Jacoby:2006}; (11) \citet{Vleeschower:2026}}
\end{table*}

The DNSs in our sample show a scarcity of systems at intermediate eccentricities, as also presented in Fig.~\ref{fig:two_panel_e_prob_masses}, with an absence of systems at measured eccentricities between 0.4 and 0.58, and one system (J1208--5936) having an eccentricity of ${\simeq}0.47$ after integrating its orbit back in time. This distribution therefore hints at a possible bimodality, although the current sample is too small to establish such a feature statistically (see Appendix \ref{appGapSignificance}).

We note that there are several observational biases at play that shape the observed DNS properties. Firstly, DNSs that form in tight binaries will have relatively short merger times due to the emission of GWs, limiting their observable window in pulsar surveys; very short-period DNSs, below about 15 minutes, are difficult to extract due to the rapid change in the observed pulsation period caused by pulsar acceleration (e.g., \citealt{RomeroShaw:2020}). Therefore, there is an observational bias against short periods and against (very) high eccentricities. Secondly, the DNSs in our sample are observed because one of the NSs is visible as a pulsar, meaning that there might be an observational bias in the orbital properties if there is a causal relationship between the binary evolution and the pulsar lifetime. Lastly, if the binary receives a significant kick and therefore a high systemic velocity, its resulting Galactic orbit will typically be at larger distances from the solar neighborhood, making it less likely to be observed \citep{Disberg:2024a,Disberg:2025}. 

% ==========================================================
\section{Exploring Bimodality in DNS Eccentricities}
\label{sec:BimodalEccentricity}
% ==========================================================

In this section, we show that an eccentricity bimodality can arise naturally in a model where the progenitor mass-NS mass relation is discontinuous and the second-born NS receives only a small natal kick. If the natal kick velocity of the second SN in the DNS system is very low, then the post-SN orbits are mostly determined by the Blaauw kicks. A break in the remnant mass function, i.e., non-monotonicity as a function of the progenitor mass of the second SN progenitor, would produce a bimodal ejecta-mass distribution and therefore a bimodal Blaauw kick distribution, which would translate to an apparent gap in eccentricity.  To demonstrate that this can produce an eccentricity distribution which resembles the current Galactic DNS sample, we employ a population synthesis simulation adopting the \cite{MandelMueller:2020} SN recipes with small USSN kicks. 

% ==========================================================
\subsection{Non-monotonicity in SN Remnant Mass Function}
\label{subsec:MassBreak}
% ==========================================================

There is no consensus in the literature regarding the relationship between progenitor and NS masses, with different studies finding different trends \citep[e.g.][]{Fryer:2012,Ugliano:2012,Nakamura:2015,Mueller:2016,Ertl:2016,Sukhbold:2016,PattonSukhbold:2020,Shishkin:2023,Burrows:2024}. We consider a physically plausible break in the mass of the newly formed NS as a function of the progenitor mass, as found in several previous studies (e.g., \citealt{MandelMueller:2020}; \citealt{Schneider:2023}; \citealt{Maltsev:2025}). 

In lower-mass core-collapse SN progenitors, carbon typically burns convectively at the center, while in more massive stars it tends to burn radiatively \citep[e.g.][]{Arnett:1972,Timmes1996,Brown:2001,Sukhbold:2020,Laplace:2025}, producing different silicon-core masses and density profiles prior to collapse. The resulting structural differences can shift the shock revival time and explosion energy, leading to different amounts of mass accreted onto the proto-NS prior to explosion and hence to a discontinuity or branching in the resulting NS mass function. Furthermore, a discontinuity can arise as a consequence of discrete composition-shell interfaces in the pre-SN structure. Density and entropy typically change abruptly across major shell boundaries (e.g., Si/O or O/C), and when such an interface reaches the shock, the associated drop in accretion ram pressure can facilitate shock acceleration (e.g., \citealt{Summa:2016}; \citealt{YamamotoYamada2016}). Therefore, shell boundaries act as natural anchors for the effective mass cut, separating ejected and retained material. Small variations in progenitor mass can shift which interface is dynamically most important for triggering shock runaway, so the mass cut can move from one boundary to another, producing a break in the resulting NS mass distribution even for similar core masses.

As a result, several studies find non-monotonicity in the remnant mass function. In the \citet{MandelMueller:2020} prescription, the core mass exhibits a sharp change at a CO core mass of $3M_\odot$, associated with a transition in the structure of the final burning shells. \citet{MandelMueller:2020} account for this behavior by introducing a break in the remnant-mass prescription at this core mass. \citet{Nair:2025}, however, find that this feature could lead to an overproduction of high-mass DNSs (as will also be discussion in Section~\ref{subsec:MainResults}. The models of \citet{Schneider:2020} and \citet{Maltsev:2025} have two different windows of NS formation in terms of the carbon-oxygen (CO) core mass, providing a natural bifurcation, but the more massive branch would require progenitors with CO core masses $\gtrsim 7$ M$_\odot$ at the time of the second SN, leading to binaries that would almost always be disrupted by \citet{Blaauw:1961} kicks. The same alternation between NS and black hole forming regimes may also create distinct black hole mass branches, potentially explaining (apparent) features in the mass distribution of binary black holes observed through GWs \citep[e.g.,][]{DisbergNelemans:2023,Schneider:2023,Laplace:2025,Willcox:2025}; The \citet{BoccioliFragione:2024} SN models, indicate a non-monotonicity between the mass of the silicon-oxygen core and the remnant NS mass, but the dependence on the CO core in these models is non-trivial due to the complex interplay between compactness and explosion energy \citep{Boccioli:2025}. 

Taken together, the physical mechanisms discussed above, along with the existence of breaks or bifurcations as found in several previous studies, motivate us to adopt a break in the NS remnant mass function, and use the \citet{MandelMueller:2020} prescription as an illustrative case. 

% ==========================================================
\subsection{Illustration with Population Synthesis}
\label{subsec:PopSynthMethod}
% ==========================================================

We illustrate our argument by adopting the \citealt{MandelMueller:2020} SN prescription within the population synthesis code \textsc{compas} (\citealt{Stevenson:2017}; \citealt{VignaGomez:2018}; \citealt{COMPAS:2021}; \citealt{COMPAS:2025}), which features a mass break between the CO core and the resulting NS mass. It connects the core mass of the exploding star and the mass of the resulting NS or black hole remnant, and stochastically assigns their natal kicks based on the CO core mass ejected in the explosion and the remnant mass. We use $630 \rm\: km \:s^{-1}$ for the NS kick scaling parameter in this model and take a NS kick distribution width of $\sigma _{\rm kick,NS}=0.45$ (\citealt{Kapil:2022}, as amended by \citealt{Disberg:2026,Disberg:2026b}). We treat USSNe, through which most of the second NSs in DNS systems are formed in our simulations, separately, and assume they have lower kick velocities. These SNe are defined as events in which the progenitor has been stripped down to its CO core prior to collapse, typically following case BB mass transfer. For illustrative purposes, we run a model where USSN kick velocities are set to zero. Hereafter, we will refer to this model as $\rm M\&M{,}0$.  

\textsc{compas} evolves each star using the analytic fitting formulae of \citealt{Hurley:2000}, calibrated to the stellar models of \cite{Pols:1998}, and models binary interactions following recipes based on \cite{Hurley:2002} but modified as described in the references above. For $10^6$ binaries, we draw the zero-age MS mass of the primary (initially more massive star) from the Kroupa initial mass function \citep{Kroupa:2001}, in the range $5 M_{\rm \odot}\le M_{\rm primary} \le 100 M_{\rm \odot}$, and the zero-age MS mass of the secondary companion $M_{\rm secondary}$ according to a flat distribution in mass ratio $ 0.1 \le q\equiv M_{\rm secondary}/M_{\rm primary} \le 1$ \citep{Sana:2012}. We sample the initial orbital separation of circular binaries from a flat-in-the-log distribution in the range  $0.1 \le a / \rm au \le 1000$ \citep{Opik1924}. All the stars in our population model have solar metallicity $Z_{\rm \odot}=0.0142$ \citep{Asplund:2009}. Running a model with a lognormal metallicity distribution between metallicities of $0.001$ and $0.03$, which favors low metallicities, has a negligible impact on the trends we find. This is in agreement with recent studies that showed metallicity does not play a significant role for DNS formation and properties (e.g, \citealt{VignaGomez:2018}; \citealt{vanSon:2024}). Therefore, even though Galactic DNSs were formed at a range of metallicities, we use solar metallicity as a representative value. All other model parameters are taken from the default \textsc{compas} settings.

% ==========================================================
\subsection{Comparison with Galactic DNS Observations}
\label{subsec:MainResults}
% ==========================================================

\begin{figure*}
    \centering
    \includegraphics[width=\linewidth]{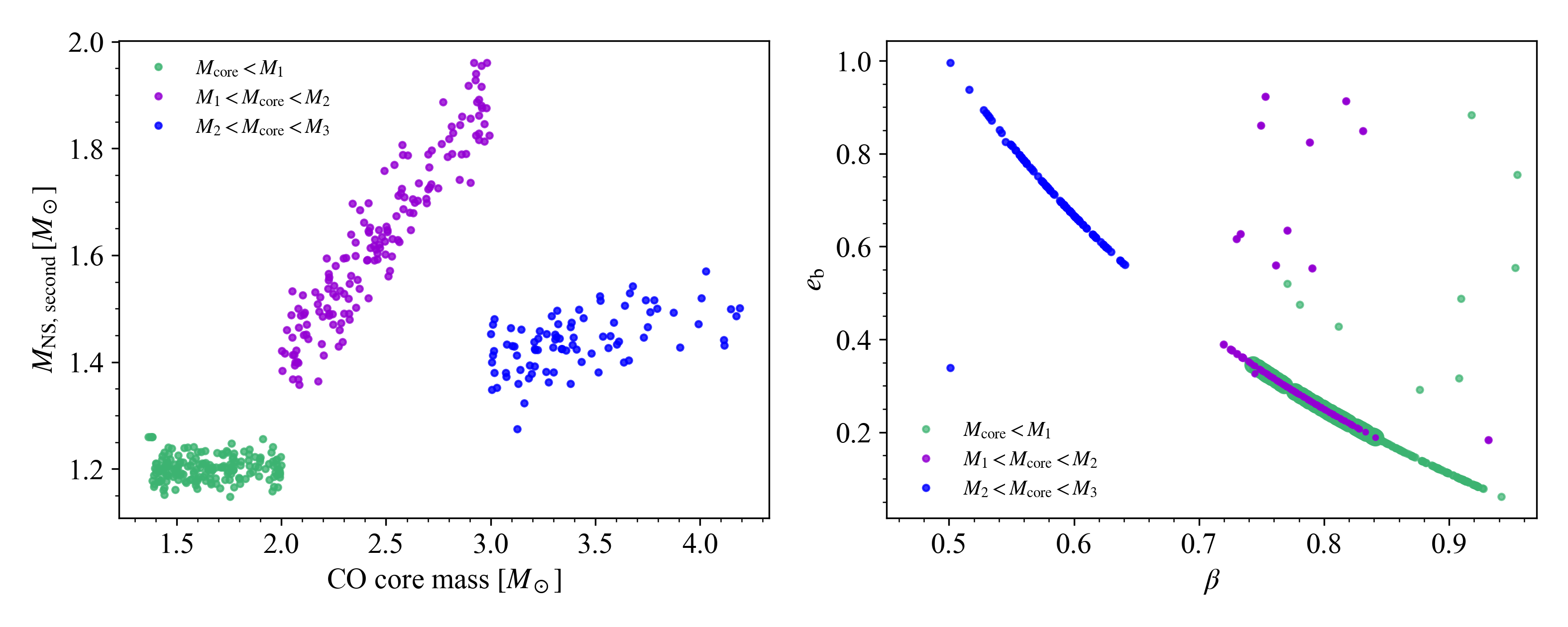}
    \caption{Properties of DNSs evolved with the \cite{MandelMueller:2020} prescription for the remnant mass and kick distribution, assuming the USSN kicks equal zero;  $\rm M\&M{,}0$. The different colors correspond to different CO core mass regimes in the \cite{MandelMueller:2020} formalism, where $M_{\rm 1}=2 M_{\rm \odot}$, $M_{\rm 2}=3 M_{\rm \odot}$, and $M_{\rm 3}=7 M_{\rm \odot}$. Left panel: mass of the NS formed in the second SN explosion vs the CO core mass, which is equivalent to the mass of the SN progenitor in an USSN.
     Right panel: eccentricity just after DNS formation vs fraction $\beta$ of total mass retained in the binary during the second SN explosion.} 
    \label{fig:two_panels_COcore_vs_MNS2_and_beta_vs_e}
\end{figure*}

The left panel of Fig.~\ref{fig:two_panels_COcore_vs_MNS2_and_beta_vs_e} shows the second NS mass at DNS formation as a function of the CO core mass (equal to the total mass for USSN progenitors) for the three different core mass regimes in the \cite{MandelMueller:2020} SN prescription implemented in our population synthesis simulation. For $M_{\rm core}<M_{\rm 1}$, where $M_{\rm 1}=2M_{\rm \odot}$, the central remnant is always an NS (green dots); if  $M_{\rm 1}\leq M_{\rm core}<M_{\rm 3}$, where $M_{\rm 3}=7M_{\rm \odot}$, the compact remnant has a probability of being an NS or a black hole, which depends on the CO core mass. The purple and blue dots show remnants that resulted in NSs. Despite the stochasticity in the recipe that leads to a scatter in the results, we can see there is a pronounced jump in the remnant mass at a CO core mass of $M_{\rm 2}=3M_{\rm \odot}$ (between the purple and blue dots).

Consider a binary with a secondary pre-SN progenitor just below this threshold mass, an ultra-stripped $2.99 M_\odot$ CO core.  In most cases, the primary is a $\sim 1.26$ M$_\odot$ NS formed through an electron-capture SN. At collapse, the secondary loses $\lesssim 1.2 $ M$_\odot$. The retained total mass fraction is $\beta \gtrsim 0.72$, producing a binary with an eccentricity of $\lesssim 0.4$ in the absence of a natal kick (right panel of Fig.~\ref{fig:two_panels_COcore_vs_MNS2_and_beta_vs_e}).  On the other hand, if the secondary progenitor had a mass of 3.01 M$_\odot$, just above the threshold mass, it would lose $\gtrsim 1.5$ M$_\odot$ during the SN, resulting in $\beta \lesssim 0.64$ and leaves behind a DNS with an eccentricity of $\gtrsim 0.55$. This is what gives rise to an eccentricity gap, as shown in the right panel of Fig.~\ref{fig:two_panels_COcore_vs_MNS2_and_beta_vs_e}. The few systems that deviate from the main trend correspond to cases in which the second SN is not classified as an USSN event, and therefore receives a non-zero natal kick.

\begin{figure*}
    \centering
    \includegraphics[width=\linewidth]{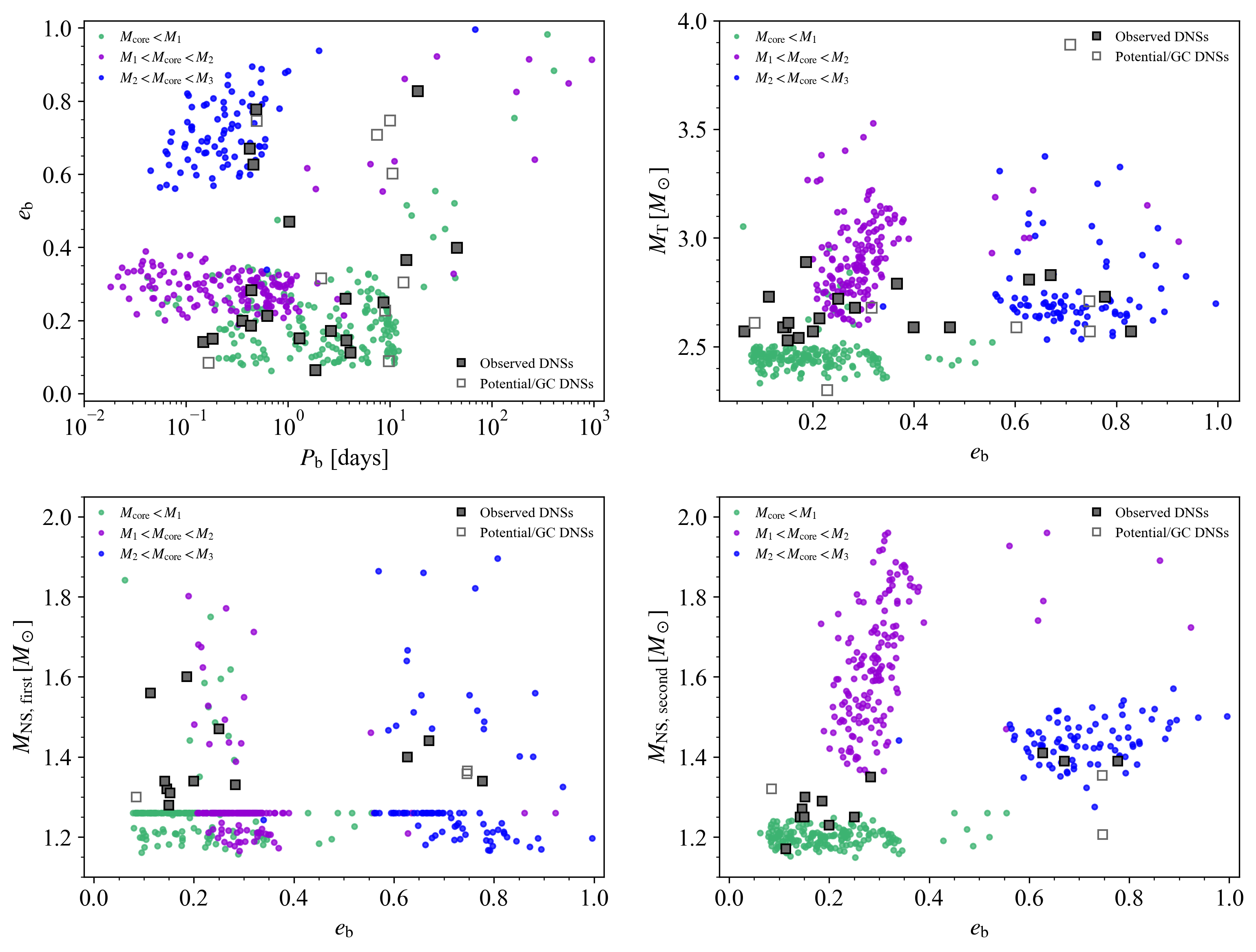}
    \caption{Observables of Galactic DNSs. Dots show the results of the $\rm M\&M{,}0$ model, with colors defined as in Fig.~\ref{fig:two_panels_COcore_vs_MNS2_and_beta_vs_e}. The gray squares are observations of Galactic DNSs taken from the ATNF Pulsar Catalog and integrated back in time to DNS formation for comparison. Empty squares are potential DNSs, or DNSs that are associated with GCs (as explained in Section~\ref{sec:observations}). Top left panel: orbital period -- eccentricity distribution immediately after the second SN explosion. Top right panel: total mass of the DNS system plotted against  eccentricity at the same time (observed DNSs with no total mass measurement are excluded). Bottom left panel: first-born NS mass vs eccentricity of the DNS binary at birth. Bottom right panel: same as bottom left panel but for the second-born NS mass. } 
    \label{fig:two_panel_e_prob_masses}
\end{figure*}

In the top left panel of Fig.~\ref{fig:two_panel_e_prob_masses} we present the orbital period -- eccentricity distribution of the $\rm M\&M{,}0$ simulated DNS population colored according to the different CO core mass regimes of the \cite{MandelMueller:2020} prescription, and compare it to observations of confirmed DNSs (filled gray squares), as well as potential DNSs and DNSs associated with GCs (open gray squares, see Section~\ref{sec:observations} for details). We find that the eccentricity gap that arises from the combination of the \cite{MandelMueller:2020} remnant mass break and lack of USSN kick lies at approximately the same eccentricities as the sparsely populated region of the the back-integrated current Galactic DNS sample. Models with significant natal kicks or a monotonic NS mass prescription do not predict a DNS eccentricity gap, as can be seen in Fig.~\ref{fig:eP_and_kick_hist_2panel} in Appendix~\ref{appMoreModels}. Gradually increasing the USSN kick results in the gap's eventual disappearance (Fig.~\ref{fig:Different_USSN_kicks} in Appendix~\ref{appMoreModels}). The absence of observed systems at short orbital periods is consistent with the previously discussed selection effects. We note that our population model underproduces the observed wide systems at high eccentricities. This discrepancy may reflect uncertainties in the late binary evolution preceding the second SN, including the fraction of systems that enter CEE or undergo case BB mass transfer, the evolutionary stage at which these interactions begin, and the resulting degree of orbital tightening, as well as selection effects that favor longer-period systems, which generally have longer GW merger times. For example, \citet{Disberg:2026b} model DNS formation and weigh the simulated binaries by a factor that accounts for their lifetime, finding a significantly larger relative contribution for these long-period systems.
 
In the top right panel of Fig.~\ref{fig:two_panel_e_prob_masses}, we show the distribution of total binary mass as a function of eccentricity in our population synthesis model, and compare it to the observed mass distribution. We focus on total binary masses inferred from pulsar timing, since individual NS masses are typically only well constrained if one NS is close to the line of sight to the pulsar, allowing the Shapiro delay to be measured, and are otherwise highly uncertain. Our simulated distribution agrees reasonably well with the observed DNS masses in some eccentricity regimes, but shows discrepancies in others. 

For binaries with low DNS eccentricities at birth ($e_{\rm b} \lesssim 0.2$), which are associated with small CO core progenitors in the \citealt{MandelMueller:2020} model (green dots), our simulations predict slightly lower masses than observed. This discrepancy may partly result from our Eddington-limited treatment of mass accretion onto the NS, which yields negligible mass growth. Observations of ultra luminous X-ray sources show that the accretion rate of NSs can surpass the Eddington limit by several orders of magnitude (e.g., \citealt{Israel:2017}; \citealt{Kaaret:2017}). Additional evidence for high accretion rates and mass gain could also be inferred from the observed spin periods of recycled pulsars, as in the case of pulsar -- white dwarf binaries (e.g., \citealt{Tauris:2012}, though in DNS systems the recycling is generally milder and might therefore indicate a more moderate mass gain, \citealt{Tauris:2017}). Mass accretion during the CEE episode, which we do not model, may further contribute to such mass growth; however, its impact on the NS mass is expected to be small (see e.g., \citealt{MacLeodRamirezRuiz:2015}; \citealt{Grichener:2023}). Overall, accounting for modest NS mass gain due to the aforementioned effects could reduce the discrepancy between the simulated and observed masses in this low-eccentricity regime.

\begin{figure}
    \centering
    \includegraphics[width=\linewidth]{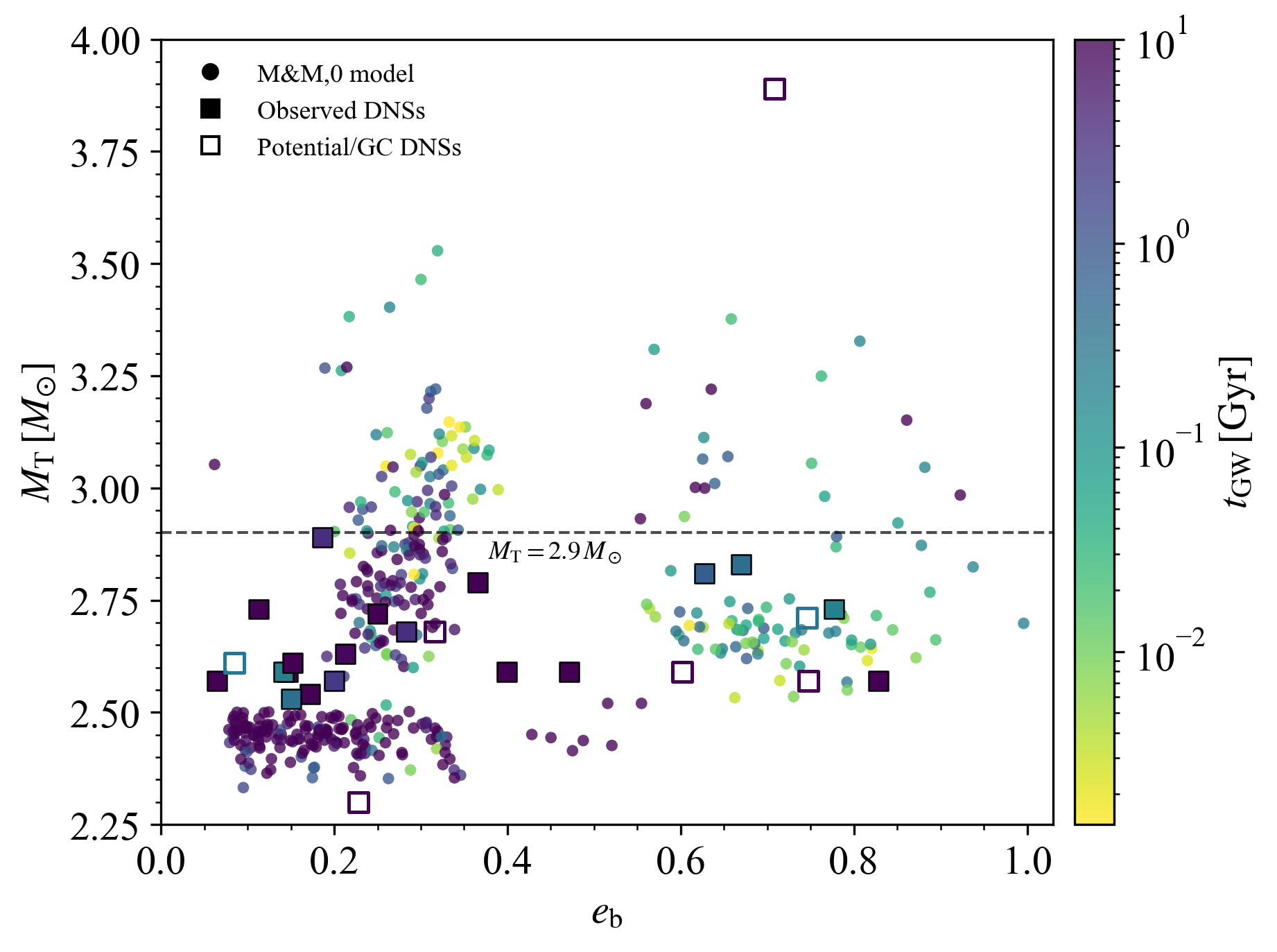} %\linewidth
    \caption{Same as the top right panel of Fig.~\ref{fig:two_panel_e_prob_masses}, but with population data and observations colored by the GW merger time, $t_{\rm GW}$. The color scale is limited to $10 \rm \; Gyr$, which is a proxy for the age of the Galactic thin disc \citep{delPeloso2005}. Systems with longer merger times are assigned the same color as $10 \rm \; Gyr$. The dashed grey line corresponds to a total binary mass of $M_{\rm T}=2.9M_{\rm \odot}$, above which observed systems are rare but our models still predict binaries. }
    \label{fig:e_Mtotal_density_plot}
\end{figure}

At the highest birth eccentricities below the currently sparsely-populated area in eccentricity observations ($0.25\lesssim e_{\rm b} \lesssim 0.4$), our model reproduces the observed mass range, but also predicts a population with higher total binary masses than currently detected, as also previously pointed out by \cite{Nair:2025}. The bottom panels of Fig.~\ref{fig:two_panel_e_prob_masses} show that this discrepancy in the total masses is primarily driven by the higher masses of the second-born NSs, which form from progenitors with CO core masses just below $3M_{\rm \odot}$ in the \citealt{MandelMueller:2020} model (purple dots). These DNSs tend to have short orbital periods (top left panel of Fig.~\ref{fig:two_panel_e_prob_masses}), implying that they will merge more rapidly due to GW emission, as shown in Fig.~\ref{fig:e_Mtotal_density_plot}, lowering the chances of detection through pulsar timing. We note, however, that a few high-eccentricity Galactic DNSs ($e_{\rm b} \gtrsim 0.6$) have been detected, despite also having short GW merger times. Furthermore, we do not account for radio lifetimes, which affect how long the NSs can be observed as pulsars and remain highly uncertain due to the poorly constrained magnetic fields, birth spins, and possible recycling histories. We discuss alternative explanations for the non-detection of massive DNSs in Section~\ref{sec:discussion}.

The bottom left panel of Fig.~\ref{fig:two_panel_e_prob_masses} shows a concentration of first-born NS masses at $1.26M_{\rm \odot}$. This arises since in our models many of these NSs are formed through electron-capture SNe, and are consequently assigned a remnant of this mass. Their relatively small CO-core ejecta naturally lead to lower kicks, since the kick magnitude scales with the ejected CO-core mass relative to the remnant mass \citep{MandelMueller:2020}. This increases the probability that the binary remains bound after the first SN and would therefore be able to continue its evolution towards a DNS system. The narrow concentration is further reinforced by the limited mass growth of the first-born NSs in our simulations, where accretion is Eddington-limited, including during case BB mass transfer (cf.~\citealt{Tauris:2015}), and mass growth during CEE is neglected (cf.~\citealt{MacLeodRamirezRuiz:2015}). Accounting for a range of ECSN remnant masses and additional accretion during binary evolution could broaden and shift this concentration towards higher masses, potentially improving agreement with observations (e.g., \citep{Mandel:2020}). Nevertheless, a preference for relatively low first-born NS masses may persist, given the increased likelihood of binary survival following the weaker natal kicks associated with electron-capture SNe.

An additional noteworthy result is the qualitative agreement of the observed second-born NS masses and eccentricities with the predictions of the $\rm M\&M{,}0$ model (bottom-right panel of Fig.~\ref{fig:two_panel_e_prob_masses}), particularly for the low- and high-core-mass regimes (green and blue points). For a circular pre-SN orbit and negligible natal kicks, the post-SN eccentricity is determined by the fraction of mass retained in the binary (see right panel of Fig.~\ref{fig:two_panels_COcore_vs_MNS2_and_beta_vs_e}). Combined with the adopted remnant-mass prescription, this provides a direct connection between the second-born NS mass and orbital eccentricity that is relatively insensitive to uncertainties in the preceding binary evolution. Although current observations do not probe the predicted high-NS-mass portion of the intermediate core-mass regime (purple points), future measurements of second-born NS masses could provide further constraints on the remnant-mass relation and test the predictions of our model.

% ==========================================================
\section{Constraints on Second SN Natal Kick}
\label{sec:SecondNSkick}
% ==========================================================

While velocity measurements of isolated pulsars suggest typical kick velocities of hundreds of $\rm km/s$ \citep[e.g.][]{IgoshevVerbunt:2017,Igoshev:2020,DisbergMandel:2025}, there is growing observation-driven evidence that SN kicks in binary systems that evolve to DNSs might be fairly small. When the first SN occurs, for instance, the binary must be in a wide orbit according to most models to allow successful envelope ejection during the subsequent CEE phase between the NS and its companion. To avoid disruption in such a wide orbit, the first SN kick has to be weak, favoring electron-capture SNe in DNS progenitors (e.g., \citealt{GiacobboMapelli:2018}; \citealt{Willcox:2021}). The second SN in most DNS systems is believed to impart relatively low kick velocities on the newly born NS \citep[e.g.,][]{Dewi:2005,vanDenHeuvel:2007,Andrews:2015,Chruslinska:2017,BeniaminiPiran:2016,ShaoLi:2018}, which hints towards a USSN. They mostly occur when a stripped helium star retains only a small helium envelope, potentially after case BB mass transfer removes most of the envelope from the helium-star progenitor (e.g., \citealt{Tauris:2013}; \citealt{Tauris:2015}; \citealt{Tauris:2017}; \citealt{Moriya:2017}; \citealt{De:2018}; \citealt{Chattaraj:2025}). At the population level, \citet{Disberg:2026b} show that the DNS systemic-kick estimates of \citet{Disberg:2024b} are consistent with small natal kicks. In particular, they identify a correlation between eccentricity and systemic kick that is expected when Blaauw kicks dominate the orbital response to the second SN, and that the observed DNS population is broadly consistent with this prediction.

Pulsar timing observations of DNSs can provide a powerful framework for placing better constraints on their kick magnitudes through spin--orbit misalignments. Consider the low-eccentricity double pulsar J0737-3039 ($e_{\rm m} = 0.088$; integrated back to $e_{\rm b}=0.147$ at DNS formation) as an example. The first NS is a recycled pulsar with a 22.7 ms spin period \citep{Burgay:2003jj}. Since it was recycled by accretion from its companion, we can expect its spin axis to be aligned with the binary's angular momentum axis before the second SN. Currently, it is observed to be misaligned by less than $3.2^\circ$  from the binary's orbital axis \citep{Ferdman:2013}. This constrains the component of the natal kick perpendicular to the orbital plane to be $\lesssim 35$ km s$^{-1}$, much smaller than the typical value for regular core-collapse SNe, and within the previously estimated natal kick range of $5-120 \rm \; km \; s^{-1}$ for PSR J0737-3039 \citep{Wong2010}. Therefore, this extremely well measured double pulsar provides strong constraints on the out-of-plane kick component of the second-born NS, and appears to be consistent with relatively low asymmetric natal kicks and \citet{Blaauw:1961} kicks from symmetric mass loss. Moreover, the relatively small systemic kick estimate of \citet{Disberg:2024b} for J0737-3039 is also consistent with low natal kicks \citep{Disberg:2026b}.

If the eccentricity at DNS formation, $e_{\rm b}=0.147$, is entirely due to symmetric mass loss with no natal kick, then $e_{\rm b}=(1-\beta)/\beta$, where $\beta$ is the fraction of the total binary mass retained after the explosion. This gives $\beta\simeq 0.87$ (see right panel of Fig. \ref{fig:two_panels_COcore_vs_MNS2_and_beta_vs_e}), corresponding to a total instantaneous mass loss of $\simeq 0.38~M_\odot$ that formed the currently $2.59~M_\odot$ binary. However, part of this mass loss is expected to be carried away by neutrinos during core collapse. Using $\Delta M_\nu \simeq 0.075 \left(M_{\rm NS}/M_\odot\right)^2M_\odot$ \citep[][and references therein]{Fryer:2012}, the baryonic--gravitational mass difference for the second-born NS is $\simeq 0.12~M_\odot$, although the exact value depends on the NS equation of state. Thus, in a purely Blaauw-kick interpretation, only $\simeq 0.26~M_\odot$ of baryonic material was promptly ejected during the second SN. Conversely, if the only mass lost in the supernova was in neutrinos, an in-plane natal kick of order $\lesssim 30 \; {\rm km\ s^{-1}}$ could account for the remaining eccentricity. While some fine tuning of the \citet{Blaauw:1961} mass loss and natal-kick direction could allow slightly larger values, J0737-3039 is consistent with very little prompt mass ejection and a very small natal kick in the second SN.

Using a similar simple spin-orbit tilt argument for the DNS system PSR B1534+12, which has a spin--orbit misalignment of $25 \pm4^\circ$, however, results in a perpendicular natal kick component of $\simeq150- 220 \rm \; km \; s^{-1}$. This is broadly consistent with the larger kick magnitude inferred by \cite{Thorsett2005}, $170-290 \rm \; km \; s^{-1}$, from a detailed reconstruction combining the misalignment with orbital and kinematic constraints. Nevertheless, PSR B1534+12 has a relatively low eccentricity, with $e_{\rm m}=0.274$ and $e_{\rm b}=0.283$. For illustration, assuming a second SN progenitor mass of $\sim3M_{\rm \odot}$ (e.g., \cite{Thorsett2005}), symmetric mass loss alone would give an eccentricity of $\sim 0.6$, larger than observed and within the high-eccentricity branch suggested by the current Galactic DNS sample. This suggests that a natal kick component with a direction roughly opposed to the pre-SN orbital motion could have partially counteracted the eccentricity increase due to mass loss, while an out-of-plane component still produced the observed spin-orbit misalignment. Therefore, B1534+12 is in clear tension with the simplifying assumption that all second SNe in close DNSs occur with negligible natal kicks. At the same time, it illustrates that the mapping between kick magnitude and post-SN eccentricity can also be direction-dependent.

For PSR B1913+16, the relatively large spin--orbit misalignment of  $18 \pm6^\circ$ (e.g., \citealt{Willems:2004}; roughly based on \citealt{Kramer:1998}) implies a significant kick component perpendicular to the pre-SN orbital plane. Applying the same tilt argument as above gives a perpendicular kick of order $100-200 \rm \; km \; s^{-1}$, consistent with detailed reconstructions (e.g., \citealt{Wong:2010}; \citealt{Tauris:2017}). In this case, its back-integrated eccentricity, $e_{\rm b}=0.670$, could be produced by symmetric mass loss alone for an illustrative second-SN progenitor of $\simeq3.3\,M_\odot$ (e.g., \citealt{Wong2010}). Thus, unlike B1534+12, the high eccentricity of B1913+16 does not itself require a large natal kick, while its spin--orbit misalignment might be an indication for such a kick.

% ==========================================================
\section{Discussion}
\label{sec:discussion}
% ==========================================================

Our results highlight how small natal kicks during the second SN can allow the non-monotonic SN remnant-mass relation to play a dominant role in shaping the eccentricity distribution of DNSs. Nevertheless, several uncertainties remain in the predicted properties of these systems.

The total mass of Galactic DNSs observed with pulsar timing at intermediate eccentricities, for instance, is generally lower than our population model predictions (top right panel of Fig.~\ref{fig:two_panel_e_prob_masses}). However, we note that the detected DNS merger GW190425 has a total mass of about $3.4 M_{\rm \odot}$ \citep{Abbott:2020}, coinciding with our higher predicted range. This could be consistent with the possibility that some high-mass DNSs are formed with short orbital periods and rapid merger timescales (Fig.~\ref{fig:e_Mtotal_density_plot}). Such systems would be detectable, or even favored, in GW observations, but could largely be missed by radio pulsar searches, biasing the observed radio population against higher-mass systems.

Another possibility is that the USSN in which the second NS is born is a very weak explosion, in which not all of the  mass is ejected promptly. Since only mass loss that occurs on a timescale shorter than the orbital timescale can be regarded as instantaneous and produces a \citet{Blaauw:1961} kick, the mass and eccentricity of the DNS at birth could be consistent with a scenario in which $\sim 1$ M$_\odot$ is ejected at velocities faster than orbital, causing an eccentricity change, while the remaining $\sim 0.5$ M$_\odot$ are expelled slowly relative to the orbital velocity, with little influence on the eccentricity. It is also possible that some mass is lost through pulsations prior to the SN, again weakening the Blaauw kick while also reducing the remnant mass \citep{Woosley:2019}.

Alternatively, an eccentricity bimodality could result from partial stripping of the helium-star progenitors. The \textsc{compas} models used here assume complete removal of the helium envelope from all stripped helium stars that expand after the helium MS and undergo case BB mass transfer. Detailed models by \citet{Tauris:2015}, however, suggest that more massive helium donors, which expand less during the helium Hertzsprung gap and evolve on shorter timescales, may retain part of their envelopes prior to the SN explosion. Furthermore, some helium stars may expand less than predicted by \textsc{compas} \citep{Woosley:2019}, avoiding case BB mass transfer and ultra-stripping altogether. This could create a natural bifurcation in the progenitor masses before the second SN, and consequently in DNS eccentricities, without requiring a break in the SN remnant mass prescription.

The small number of DNSs with measured spin--orbit misalignment angles and well-constrained systemic velocities, together with systemic model uncertainties, does not yet allow the second-SN natal-kick distribution to be determined robustly. Our illustrative $\rm M\&M{,}0$ model assumes negligible USSN kicks in order to isolate the effect of the remnant-mass break, and shows that the main orbital properties of the observed DNS population can be reproduced without invoking substantial natal kicks. This assumption may be in some tension with inferences for individual systems, for which spin--orbit misalignments and kinematic reconstructions have been interpreted as evidence for relatively large natal kicks (Section~\ref{sec:SecondNSkick}). Given these uncertainties in the misalignment measurements and the small number of systems with useful constraints, a kick distribution dominated by weak kicks with a subdominant higher-kick component \citep[e.g.,][]{Podsiadlowski:2004,BeniaminiPiran:2016,Kruckow:2018} cannot be ruled out. Large natal kicks, however, would affect the orbital periods, eccentricities, and systemic velocities in non-trivial ways, and accommodating such a component may require simultaneous changes to the assumed mass-loss prescription in order to remain consistent with the observations. In the present work, we therefore use the \citet{MandelMueller:2020} remnant-mass and associated mass-loss prescription, as it is motivated by supernova simulations developed independently of the Galactic DNS observations.

Galactic DNSs are not the only binary systems in which distinct eccentricity populations have been suggested. \citet{Knigge:2011}, for instance, found evidence for two populations of Be X-ray binaries with different spin, orbital-period, and eccentricity distributions.  More recently,  \citet{Valli:2025} confirmed a clear eccentricity bifurcation among Galactic Be X-ray binaries.  They find that a conventional two-component isotropic natal-kick distribution does not reproduce this separation, instead populating the intermediate region. However, the high-eccentricity branch exhibits a period-eccentricity correlation which appears to be difficult to explain with a \citet{Blaauw:1961} kick alone. Indeed, \citet{Disberg:2026b} argue that the pre-SN orbital eccentricity may contribute substantially to the eccentricities observed in the Be X-ray binary sample of \citet{Valli:2025}, with sufficiently weak natal kicks allowing this eccentricity to be preserved through the SN \citep[see also][]{Shariat_2026}. In the \citet{Disberg:2026b} model, the NS populations in both Be X-ray binaries and DNSs receive negligible natal kicks, potentially pointing to a dichotomy between NSs formed in binaries with high-mass companions and those formed with low-mass companions or in isolation. Pulsar--white dwarf binaries may also have a less pronounced eccentricity gap at slightly higher eccentricities of $\sim 0.5-0.6$. However, eccentric pulsar -- white dwarf binaries tend to be associated with GCs \citep{Disberg:2026b}, where dynamical interactions could provide an explanation for their eccentricities.

% ==========================================================
\section{Summary}
\label{sec:summary}
% ==========================================================

In this study, we showed that a non-monotonic relationship between progenitor mass and NS mass, combined with negligible natal kicks in the second SN during DNS formation, can naturally produce a bimodal Blaauw kick distribution and, consequently, a bimodality in DNS eccentricities. In particular, we curated a sample of observed DNSs (Section~\ref{sec:observations}), and used the population synthesis code \textsc{compas} to illustrate this model (Section~\ref{sec:BimodalEccentricity}).  

There are several reasons to expect the relationship between the progenitor mass and NS mass to be non-monotonic (Section~\ref{subsec:MassBreak}). Variations in pre-SN core structure, especially around the transition from convective to radiative carbon burning, can produce sharp changes in the collapse and explosion dynamics, leading to different amounts of fallback or accretion onto the proto-NS. In addition, abrupt composition-shell interfaces can shift the effective mass cut between ejected and retained material, naturally giving rise to a discontinuity or branching in the NS mass distribution. Therefore, we adopted the \cite{MandelMueller:2020} prescription that features such a break (Section~\ref{subsec:PopSynthMethod}). 

We found that the resulting DNS population broadly resembles the observed orbital period---eccentricity distribution, including a deficit at intermediate eccentricities, where the current Galactic DNS sample is sparsely populated. The observed systems also largely fall within the regions populated by our model in the second-born NS mass---eccentricity plane, particularly the predictions for low and high core masses (Section~\ref{subsec:MainResults}).
However, the current sample is too small to establish an eccentricity bimodality (Appendix \ref{appGapSignificance}) or robustly constrain the second-SN natal-kick distribution (Section~\ref{sec:SecondNSkick}). Expanding the Galactic sample with SKA \citep{Braun:2019}, improved mass measurements, and more detailed modeling of SN ejecta and kick physics will be crucial for testing our model and determining whether an eccentricity gap remains consistent with observations.

\section*{Acknowledgments}

We are grateful to Jeff Andrews, Mike Lau, Marcus Lower, Bernhard M\"{u}ller, Gijs Nelemans, Mathieu Renzo, Thomas Tauris and Reinhold Willcox for very useful comments and stimulating discussions. We thank the anonymous referee for useful suggestions that helped to improve this manuscript. AG acknowledges support from the Steward Observatory Fellowship in Theoretical and Computational Astrophysics, the IAU-Gruber Fellowship, and the CHE Fellowship. PD, RH and IM acknowledge support from the Australian Research Council (ARC) Centre of Excellence for Gravitational Wave Discovery (OzGrav), through project number CE230100016. 

% ==========================================================
\section*{Data availability}
\label{sec:DataAvailability}
% ==========================================================

In this work we used the population synthesis code \textsc{compas}, version 03.27.01. The population synthesis data generated for this project is publicly available on Zenodo at \href{https://doi.org/10.5281/zenodo.21651784}{doi:10.5281/zenodo.21651784}.

\bibliography{Mandel.bib}{}

@Article{Peters:1964,
   author = {{Peters}, P.~C.},
    title = "{Gravitational Radiation and the Motion of Two Point Masses}",
  journal = {Physical Review},
     year = 1964,
    month = nov,
   volume = 136,
    pages = {1224-1232},
      doi = {10.1103/PhysRev.136.B1224},
   adsurl = {http://adsabs.harvard.edu/abs/1964PhRv..136.1224P}
}

@Article{Burgay:2003jj,
     author    = "Burgay, Marta and others",
     title     = "An increased estimate of the merger rate of double neutron
                  stars from observations of a highly relativistic system",
     journal   = "Nature",
     volume    = "426",
     year      = "2003",
     pages     = "531-533",
     eprint    = "astro-ph/0312071",
     SLACcitation  = "%%CITATION = ASTRO-PH 0312071;%%"
}

@ARTICLE{Pfahl:2002,
   author = {{Pfahl}, E. and {Rappaport}, S. and {Podsiadlowski}, P.},
    title = "{A Comprehensive Study of Neutron Star Retention in Globular Clusters}",
  journal = {\apj},
   eprint = {arXiv:astro-ph/0106141},
     year = 2002,
    month = jul,
   volume = 573,
    pages = {283-305},
      doi = {10.1086/340494},
   adsurl = {http://adsabs.harvard.edu/abs/2002ApJ...573..283P}
}

@ARTICLE{Fryer:2012,
   author = {{Fryer}, C.~L. and {Belczynski}, K. and {Wiktorowicz}, G. and 
	{Dominik}, M. and {Kalogera}, V. and {Holz}, D.~E.},
    title = "{Compact Remnant Mass Function: Dependence on the Explosion Mechanism and Metallicity}",
  journal = {\apj},
archivePrefix = "arXiv",
   eprint = {1110.1726},
 primaryClass = "astro-ph.SR",
     year = 2012,
    month = apr,
   volume = 749,
      eid = {91},
    pages = {91},
      doi = {10.1088/0004-637X/749/1/91},
   adsurl = {http://adsabs.harvard.edu/abs/2012ApJ...749...91F}
}

@ARTICLE{Sana:2012,
   author = {{Sana}, H. and {de Mink}, S.~E. and {de Koter}, A. and {Langer}, N. and 
	{Evans}, C.~J. and {Gieles}, M. and {Gosset}, E. and {Izzard}, R.~G. and 
	{Le Bouquin}, J.-B. and {Schneider}, F.~R.~N.},
    title = "{Binary Interaction Dominates the Evolution of Massive Stars}",
  journal = {Science},
archivePrefix = "arXiv",
   eprint = {1207.6397},
 primaryClass = "astro-ph.SR",
     year = 2012,
    month = jul,
   volume = 337,
    pages = {444-},
      doi = {10.1126/science.1223344},
   adsurl = {http://adsabs.harvard.edu/abs/2012Sci...337..444S}
}

@ARTICLE{Kroupa:2001,
   author = {{Kroupa}, P.},
    title = "{On the variation of the initial mass function}",
  journal = {\mnras},
   eprint = {astro-ph/0009005},
     year = 2001,
    month = apr,
   volume = 322,
    pages = {231-246},
      doi = {10.1046/j.1365-8711.2001.04022.x},
   adsurl = {http://adsabs.harvard.edu/abs/2001MNRAS.322..231K}
}

@ARTICLE{Blaauw:1961,
   author = {{Blaauw}, A.},
    title = "{On the origin of the O- and B-type stars with high velocities (the ''run-away'' stars), and some related problems}",
  journal = {Bull.~Astron.~Inst.~Netherlands},
     year = 1961,
    month = may,
   volume = 15,
    pages = {265},
   adsurl = {http://adsabs.harvard.edu/abs/1961BAN....15..265B}
}

@ARTICLE{Asplund:2009,
   author = {{Asplund}, M. and {Grevesse}, N. and {Sauval}, A.~J. and {Scott}, P.
	},
    title = "{The Chemical Composition of the Sun}",
  journal = {\araa},
archivePrefix = "arXiv",
   eprint = {0909.0948},
 primaryClass = "astro-ph.SR",
     year = 2009,
    month = sep,
   volume = 47,
    pages = {481-522},
      doi = {10.1146/annurev.astro.46.060407.145222},
   adsurl = {http://adsabs.harvard.edu/abs/2009ARA\%26A..47..481A}
}

@ARTICLE{Hurley:2000,
   author = {{Hurley}, J.~R. and {Pols}, O.~R. and {Tout}, C.~A.},
    title = "{Comprehensive analytic formulae for stellar evolution as a function of mass and metallicity}",
  journal = {\mnras},
   eprint = {astro-ph/0001295},
     year = 2000,
    month = jul,
   volume = 315,
    pages = {543-569},
      doi = {10.1046/j.1365-8711.2000.03426.x},
   adsurl = {http://adsabs.harvard.edu/abs/2000MNRAS.315..543H}
}

@ARTICLE{Hurley:2002,
   author = {{Hurley}, J.~R. and {Tout}, C.~A. and {Pols}, O.~R.},
    title = "{Evolution of binary stars and the effect of tides on binary populations}",
  journal = {\mnras},
   eprint = {astro-ph/0201220},
     year = 2002,
    month = feb,
   volume = 329,
    pages = {897-928},
      doi = {10.1046/j.1365-8711.2002.05038.x},
   adsurl = {http://adsabs.harvard.edu/abs/2002MNRAS.329..897H}
}

@ARTICLE{Pols:1998,
   author = {{Pols}, O.~R. and {Schr{\"o}der}, K.-P. and {Hurley}, J.~R. and 
	{Tout}, C.~A. and {Eggleton}, P.~P.},
    title = "{Stellar evolution models for Z = 0.0001 to 0.03}",
  journal = {\mnras},
     year = 1998,
    month = aug,
   volume = 298,
    pages = {525-536},
      doi = {10.1046/j.1365-8711.1998.01658.x},
   adsurl = {http://adsabs.harvard.edu/abs/1998MNRAS.298..525P}
}

@ARTICLE{Stevenson:2017,
   author = {{Stevenson}, S. and {Vigna-G{\'o}mez}, A. and {Mandel}, I. and 
	{Barrett}, J.~W. and {Neijssel}, C.~J. and {Perkins}, D. and 
	{de Mink}, S.~E.},
    title = "{Formation of the first three gravitational-wave observations through isolated binary evolution}",
  journal = {Nat. Commun.},
archivePrefix = "arXiv",
   eprint = {1704.01352},
 primaryClass = "astro-ph.HE",
     year = 2017,
    month = apr,
   volume = 8,
      eid = {14906},
    pages = {14906},
      doi = {10.1038/ncomms14906},
   adsurl = {http://adsabs.harvard.edu/abs/2017NatCo...814906S}
}

@ARTICLE{MacLeodRamirezRuiz:2015,
   author = {{MacLeod}, M. and {Ramirez-Ruiz}, E.},
    title = "{On the Accretion-fed Growth of Neutron Stars during Common Envelope}",
  journal = {\apjl},
archivePrefix = "arXiv",
   eprint = {1410.5421},
 primaryClass = "astro-ph.SR",
     year = 2015,
    month = jan,
   volume = 798,
      eid = {L19},
    pages = {L19},
      doi = {10.1088/2041-8205/798/1/L19},
   adsurl = {http://adsabs.harvard.edu/abs/2015ApJ...798L..19M}
}

@ARTICLE{Tauris:2017,
   author = {{Tauris}, T.~M. and {Kramer}, M. and {Freire}, P.~C.~C. and 
	{Wex}, N. and {Janka}, H.-T. and {Langer}, N. and {Podsiadlowski}, P. and 
	{Bozzo}, E. and {Chaty}, S. and {Kruckow}, M.~U. and {van den Heuvel}, E.~P.~J. and 
	{Antoniadis}, J. and {Breton}, R.~P. and {Champion}, D.~J.},
    title = "{Formation of Double Neutron Star Systems}",
  journal = {\apj},
archivePrefix = "arXiv",
   eprint = {1706.09438},
 primaryClass = "astro-ph.HE",
     year = 2017,
    month = sep,
   volume = 846,
      eid = {170},
    pages = {170},
      doi = {10.3847/1538-4357/aa7e89},
   adsurl = {http://adsabs.harvard.edu/abs/2017ApJ...846..170T}
}

@ARTICLE{Knigge:2011,
   author = {{Knigge}, C. and {Coe}, M.~J. and {Podsiadlowski}, P.},
    title = "{Two populations of X-ray pulsars produced by two types of supernova}",
  journal = {\nat},
archivePrefix = "arXiv",
   eprint = {1111.2051},
 primaryClass = "astro-ph.SR",
     year = 2011,
    month = nov,
   volume = 479,
    pages = {372-375},
      doi = {10.1038/nature10529},
   adsurl = {http://adsabs.harvard.edu/abs/2011Natur.479..372K}
}

@ARTICLE{Kruckow:2018,
   author = {{Kruckow}, M.~U. and {Tauris}, T.~M. and {Langer}, N. and {Kramer}, M. and 
	{Izzard}, R.~G.},
    title = "{Progenitors of gravitational wave mergers: binary evolution with the stellar grid-based code COMBINE}",
  journal = {\mnras},
archivePrefix = "arXiv",
   eprint = {1801.05433},
 primaryClass = "astro-ph.SR",
     year = 2018,
    month = dec,
   volume = 481,
    pages = {1908-1949},
      doi = {10.1093/mnras/sty2190},
   adsurl = {http://adsabs.harvard.edu/abs/2018MNRAS.481.1908K}
}

@ARTICLE{VignaGomez:2018,
   author = {{Vigna-G{\'o}mez}, A. and {Neijssel}, C.~J. and {Stevenson}, S. and 
	{Barrett}, J.~W. and {Belczynski}, K. and {Justham}, S. and 
	{de Mink}, S.~E. and {M{\"u}ller}, B. and {Podsiadlowski}, P. and 
	{Renzo}, M. and {Sz{\'e}csi}, D. and {Mandel}, I.},
    title = "{On the formation history of Galactic double neutron stars}",
  journal = {\mnras},
archivePrefix = "arXiv",
   eprint = {1805.07974},
 primaryClass = "astro-ph.SR",
     year = 2018,
    month = dec,
   volume = 481,
    pages = {4009-4029},
      doi = {10.1093/mnras/sty2463},
   adsurl = {http://adsabs.harvard.edu/abs/2018MNRAS.481.4009V}
}

@ARTICLE{HulseTaylor:1975,
   author = {{Hulse}, R.~A. and {Taylor}, J.~H.},
    title = "{Discovery of a pulsar in a binary system}",
  journal = {\apjl},
     year = 1975,
    month = jan,
   volume = 195,
    pages = {L51-L53},
      doi = {10.1086/181708},
   adsurl = {http://adsabs.harvard.edu/abs/1975ApJ...195L..51H}
}

@ARTICLE{GiacobboMapelli:2018,
       author = {{Giacobbo}, Nicola and {Mapelli}, Michela},
        title = "{The progenitors of compact-object binaries: impact of metallicity, common envelope and natal kicks}",
      journal = {\mnras},
         year = "2018",
        month = "Oct",
       volume = {480},
       number = {2},
        pages = {2011-2030},
          doi = {10.1093/mnras/sty1999},
archivePrefix = {arXiv},
       eprint = {1806.00001},
 primaryClass = {astro-ph.HE},
       adsurl = {https://ui.adsabs.harvard.edu/abs/2018MNRAS.480.2011G}
}

@ARTICLE{FlanneryvdH:1975,
   author = {{Flannery}, B.~P. and {van den Heuvel}, E.~P.~J.},
    title = "{On the origin of the binary pulsar PSR 1913 + 16}",
  journal = {\aap},
     year = 1975,
    month = feb,
   volume = 39,
    pages = {61-67},
   adsurl = {http://adsabs.harvard.edu/abs/1975A%26A....39...61F}
}

@ARTICLE{Mueller:2016,
   author = {{M{\"u}ller}, B. and {Heger}, A. and {Liptai}, D. and {Cameron}, J.~B.},
    title = "{A simple approach to the supernova progenitor-explosion connection}",
  journal = {\mnras},
archivePrefix = "arXiv",
   eprint = {1602.05956},
 primaryClass = "astro-ph.SR",
     year = 2016,
    month = jul,
   volume = 460,
    pages = {742-764},
      doi = {10.1093/mnras/stw1083},
   adsurl = {http://adsabs.harvard.edu/abs/2016MNRAS.460..742M}
}

@ARTICLE{Ertl:2016,
   author = {{Ertl}, T. and {Ugliano}, M. and {Janka}, H.-T. and {Marek}, A. and 
	{Arcones}, A.},
    title = "{Erratum: Progenitor-explosion Connection and Remnant Birth Masses for Neutrino-driven Supernovae of Iron-core Progenitors (2012, ApJ, 757, 69)}",
  journal = {\apj},
archivePrefix = "arXiv",
   eprint = {1602.06327},
 primaryClass = "astro-ph.HE",
     year = 2016,
    month = apr,
   volume = 821,
      eid = {69},
    pages = {69},
      doi = {10.3847/0004-637X/821/1/69},
   adsurl = {http://adsabs.harvard.edu/abs/2016ApJ...821...69E}
}

@ARTICLE{Sukhbold:2016,
   author = {{Sukhbold}, T. and {Ertl}, T. and {Woosley}, S.~E. and {Brown}, J.~M. and 
	{Janka}, H.-T.},
    title = "{Core-collapse Supernovae from 9 to 120 Solar Masses Based on Neutrino-powered Explosions}",
  journal = {\apj},
archivePrefix = "arXiv",
   eprint = {1510.04643},
 primaryClass = "astro-ph.HE",
     year = 2016,
    month = apr,
   volume = 821,
      eid = {38},
    pages = {38},
      doi = {10.3847/0004-637X/821/1/38},
   adsurl = {http://adsabs.harvard.edu/abs/2016ApJ...821...38S}
}

@ARTICLE{Tauris:2015,
   author = {{Tauris}, T.~M. and {Langer}, N. and {Podsiadlowski}, P.},
    title = "{Ultra-stripped supernovae: progenitors and fate}",
  journal = {\mnras},
archivePrefix = "arXiv",
   eprint = {1505.00270},
 primaryClass = "astro-ph.SR",
     year = 2015,
    month = aug,
   volume = 451,
    pages = {2123-2144},
      doi = {10.1093/mnras/stv990},
   adsurl = {http://adsabs.harvard.edu/abs/2015MNRAS.451.2123T}
}

@ARTICLE{Andrews:2015,
   author = {{Andrews}, J.~J. and {Farr}, W.~M. and {Kalogera}, V. and {Willems}, B.
	},
    title = "{Evolutionary Channels for the Formation of Double Neutron Stars}",
  journal = {\apj},
archivePrefix = "arXiv",
   eprint = {1410.6797},
 primaryClass = "astro-ph.SR",
     year = 2015,
    month = mar,
   volume = 801,
      eid = {32},
    pages = {32},
      doi = {10.1088/0004-637X/801/1/32},
   adsurl = {http://adsabs.harvard.edu/abs/2015ApJ...801...32A}
}

@ARTICLE{AndrewsMandel:2019,
   author = {{Andrews}, J.~J. and {Mandel}, I.},
    title = "{Double Neutron Star Populations and Formation Channels}",
  journal = {\apjl},
archivePrefix = "arXiv",
   eprint = {1904.12745},
 primaryClass = "astro-ph.HE",
     year = 2019,
    month = jul,
   volume = 880,
      eid = {L8},
    pages = {L8},
      doi = {10.3847/2041-8213/ab2ed1},
   adsurl = {https://ui.adsabs.harvard.edu/abs/2019ApJ...880L...8A}
}

@ARTICLE{IgoshevVerbunt:2017,
   author = {{Verbunt}, F. and {Igoshev}, A. and {Cator}, E.},
    title = "{The observed velocity distribution of young pulsars}",
  journal = {\aap},
archivePrefix = "arXiv",
   eprint = {1708.08281},
 primaryClass = "astro-ph.HE",
     year = 2017,
    month = dec,
   volume = 608,
      eid = {A57},
    pages = {A57},
      doi = {10.1051/0004-6361/201731518},
   adsurl = {https://ui.adsabs.harvard.edu/abs/2017A%26A...608A..57V}
}

@ARTICLE{Ye:2019,
       author = {{Ye}, Claire S. and {Fong}, Wen-fai and {Kremer}, Kyle and
         {Rodriguez}, Carl L. and {Chatterjee}, Sourav and {Fragione}, Giacomo and
         {Rasio}, Frederic A.},
        title = "{On the Rate of Neutron Star Binary Mergers from Globular Clusters}",
      journal = {\apjl},
         year = 2020,
        month = jan,
       volume = {888},
       number = {1},
          eid = {L10},
        pages = {L10},
          doi = {10.3847/2041-8213/ab5dc5},
archivePrefix = {arXiv},
       eprint = {1910.10740},
 primaryClass = {astro-ph.HE},
       adsurl = {https://ui.adsabs.harvard.edu/abs/2020ApJ...888L..10Y}
}

@ARTICLE{Breivik:2020,
       author = {{Breivik}, Katelyn and {Coughlin}, Scott and {Zevin}, Michael and {Rodriguez}, Carl L. and {Kremer}, Kyle and {Ye}, Claire S. and {Andrews}, Jeff J. and {Kurkowski}, Michael and {Digman}, Matthew C. and {Larson}, Shane L. and {Rasio}, Frederic A.},
        title = "{COSMIC Variance in Binary Population Synthesis}",
      journal = {\apj},
         year = 2020,
        month = jul,
       volume = {898},
       number = {1},
          eid = {71},
        pages = {71},
          doi = {10.3847/1538-4357/ab9d85},
archivePrefix = {arXiv},
       eprint = {1911.00903},
 primaryClass = {astro-ph.HE},
       adsurl = {https://ui.adsabs.harvard.edu/abs/2020ApJ...898...71B}
}

@ARTICLE{Ugliano:2012,
       author = {{Ugliano}, Marcella and {Janka}, Hans-Thomas and {Marek}, Andreas and
         {Arcones}, Almudena},
        title = "{Progenitor-explosion Connection and Remnant Birth Masses for Neutrino-driven Supernovae of Iron-core Progenitors}",
      journal = {\apj},
         year = 2012,
        month = sep,
       volume = {757},
       number = {1},
          eid = {69},
        pages = {69},
          doi = {10.1088/0004-637X/757/1/69},
archivePrefix = {arXiv},
       eprint = {1205.3657},
 primaryClass = {astro-ph.SR},
       adsurl = {https://ui.adsabs.harvard.edu/abs/2012ApJ...757...69U}
}

@ARTICLE{Nakamura:2015,
       author = {{Nakamura}, Ko and {Takiwaki}, Tomoya and {Kuroda}, Takami and
         {Kotake}, Kei},
        title = "{Systematic features of axisymmetric neutrino-driven core-collapse supernova models in multiple progenitors}",
      journal = {\pasj},
         year = 2015,
        month = dec,
       volume = {67},
       number = {6},
          eid = {107},
        pages = {107},
          doi = {10.1093/pasj/psv073},
archivePrefix = {arXiv},
       eprint = {1406.2415},
 primaryClass = {astro-ph.HE},
       adsurl = {https://ui.adsabs.harvard.edu/abs/2015PASJ...67..107N}
}

@ARTICLE{LyneLorimer:1994,
       author = {{Lyne}, A.~G. and {Lorimer}, D.~R.},
        title = "{High birth velocities of radio pulsars}",
      journal = {\nat},
         year = 1994,
        month = may,
       volume = {369},
       number = {6476},
        pages = {127-129},
          doi = {10.1038/369127a0},
       adsurl = {https://ui.adsabs.harvard.edu/abs/1994Natur.369..127L}
}

@ARTICLE{PattonSukhbold:2020,
       author = {{Patton}, Rachel A. and {Sukhbold}, Tuguldur},
        title = "{Towards a realistic explosion landscape for binary population synthesis}",
      journal = {\mnras},
         year = 2020,
        month = dec,
       volume = {499},
       number = {2},
        pages = {2803-2816},
          doi = {10.1093/mnras/staa3029},
archivePrefix = {arXiv},
       eprint = {2005.03055},
 primaryClass = {astro-ph.SR},
       adsurl = {https://ui.adsabs.harvard.edu/abs/2020MNRAS.499.2803P}
}

@ARTICLE{Podsiadlowski:2004,
       author = {{Podsiadlowski}, Ph. and {Langer}, N. and {Poelarends}, A.~J.~T. and
         {Rappaport}, S. and {Heger}, A. and {Pfahl}, E.},
        title = "{The Effects of Binary Evolution on the Dynamics of Core Collapse and Neutron Star Kicks}",
      journal = {\apj},
         year = 2004,
        month = sep,
       volume = {612},
       number = {2},
        pages = {1044-1051},
          doi = {10.1086/421713},
archivePrefix = {arXiv},
       eprint = {astro-ph/0309588},
 primaryClass = {astro-ph},
       adsurl = {https://ui.adsabs.harvard.edu/abs/2004ApJ...612.1044P}
}

@ARTICLE{BeniaminiPiran:2016,
       author = {{Beniamini}, Paz and {Piran}, Tsvi},
        title = "{Formation of double neutron star systems as implied by observations}",
      journal = {\mnras},
         year = 2016,
        month = mar,
       volume = {456},
       number = {4},
        pages = {4089-4099},
          doi = {10.1093/mnras/stv2903},
archivePrefix = {arXiv},
       eprint = {1510.03111},
 primaryClass = {astro-ph.HE},
       adsurl = {https://ui.adsabs.harvard.edu/abs/2016MNRAS.456.4089B}
}

@ARTICLE{MandelMueller:2020,
       author = {{Mandel}, Ilya and {M{\"u}ller}, Bernhard},
        title = "{Simple recipes for compact remnant masses and natal kicks}",
      journal = {\mnras},
         year = 2020,
        month = oct,
       volume = {499},
       number = {3},
        pages = {3214-3221},
          doi = {10.1093/mnras/staa3043},
archivePrefix = {arXiv},
       eprint = {2006.08360},
 primaryClass = {astro-ph.HE},
       adsurl = {https://ui.adsabs.harvard.edu/abs/2020MNRAS.499.3214M}
}

@ARTICLE{ShaoLi:2018,
       author = {{Shao}, Yong and {Li}, Xiang-Dong},
        title = "{On the Role of Supernova Kicks in the Formation of Galactic Double Neutron Star Systems}",
      journal = {\apj},
         year = 2018,
        month = nov,
       volume = {867},
       number = {2},
          eid = {124},
        pages = {124},
          doi = {10.3847/1538-4357/aae648},
archivePrefix = {arXiv},
       eprint = {1810.03324},
 primaryClass = {astro-ph.HE},
       adsurl = {https://ui.adsabs.harvard.edu/abs/2018ApJ...867..124S}
}

@ARTICLE{RomeroShaw:2020,
       author = {{Romero-Shaw}, Isobel M. and {Farrow}, Nicholas and {Stevenson}, Simon and {Thrane}, Eric and {Zhu}, Xing-Jiang},
        title = "{On the origin of GW190425}",
      journal = {\mnras},
         year = 2020,
        month = jul,
       volume = {496},
       number = {1},
        pages = {L64-L69},
          doi = {10.1093/mnrasl/slaa084},
archivePrefix = {arXiv},
       eprint = {2001.06492},
 primaryClass = {astro-ph.HE},
       adsurl = {https://ui.adsabs.harvard.edu/abs/2020MNRAS.496L..64R}
}

@ARTICLE{Chruslinska:2017,
       author = {{Chruslinska}, M. and {Belczynski}, K. and {Bulik}, T. and {Gladysz}, W.},
        title = "{Constraints on the Formation of Double Neutron Stars from the Observed Eccentricities and Current Limits on Merger Rates}",
      journal = {\actaa},
         year = 2017,
        month = mar,
       volume = {67},
       number = {1},
        pages = {37-50},
          doi = {10.32023/0001-5237/67.1.2},
archivePrefix = {arXiv},
       eprint = {1611.05366},
 primaryClass = {astro-ph.SR},
       adsurl = {https://ui.adsabs.harvard.edu/abs/2017AcA....67...37C}
}

@ARTICLE{Mandel:2020,
       author = {{Mandel}, Ilya and {M{\"u}ller}, Bernhard and {Riley}, Jeff and {de Mink}, Selma E. and {Vigna-G{\'o}mez}, Alejandro and {Chattopadhyay}, Debatri},
        title = "{Binary population synthesis with probabilistic remnant mass and kick prescriptions}",
      journal = {\mnras},
         year = 2021,
        month = jan,
       volume = {500},
       number = {1},
        pages = {1380-1384},
          doi = {10.1093/mnras/staa3390},
archivePrefix = {arXiv},
       eprint = {2007.03890},
 primaryClass = {astro-ph.HE},
       adsurl = {https://ui.adsabs.harvard.edu/abs/2021MNRAS.500.1380M}
}

@ARTICLE{Schneider:2020,
       author = {{Schneider}, F.~R.~N. and {Podsiadlowski}, Ph. and {M{\"u}ller}, B.},
        title = "{Pre-supernova evolution, compact-object masses, and explosion properties of stripped binary stars}",
      journal = {\aap},
         year = 2021,
        month = jan,
       volume = {645},
          eid = {A5},
        pages = {A5},
          doi = {10.1051/0004-6361/202039219},
archivePrefix = {arXiv},
       eprint = {2008.08599},
 primaryClass = {astro-ph.SR},
       adsurl = {https://ui.adsabs.harvard.edu/abs/2021A&A...645A...5S}
}

@ARTICLE{Willcox:2021,
       author = {{Willcox}, Reinhold and {Mandel}, Ilya and {Thrane}, Eric and {Deller}, Adam and {Stevenson}, Simon and {Vigna-G{\'o}mez}, Alejandro},
        title = "{Constraints on Weak Supernova Kicks from Observed Pulsar Velocities}",
      journal = {\apjl},
         year = 2021,
        month = oct,
       volume = {920},
       number = {2},
          eid = {L37},
        pages = {L37},
          doi = {10.3847/2041-8213/ac2cc8},
archivePrefix = {arXiv},
       eprint = {2107.04251},
 primaryClass = {astro-ph.HE},
       adsurl = {https://ui.adsabs.harvard.edu/abs/2021ApJ...920L..37W}
}

@ARTICLE{MandelBroekgaarden:2021,
       author = {{Mandel}, Ilya and {Broekgaarden}, Floor S.},
        title = "{Rates of compact object coalescences}",
      journal = {Living Reviews in Relativity},
         year = 2022,
        month = dec,
       volume = {25},
       number = {1},
          eid = {1},
        pages = {1},
          doi = {10.1007/s41114-021-00034-3},
archivePrefix = {arXiv},
       eprint = {2107.14239},
 primaryClass = {astro-ph.HE},
       adsurl = {https://ui.adsabs.harvard.edu/abs/2022LRR....25....1M}
}

@ARTICLE{COMPAS:2021,
       author = {{Team COMPAS: Riley}, Jeff and {Agrawal}, Poojan and {Barrett}, Jim W. and {Boyett}, Kristan N.~K. and {Broekgaarden}, Floor S. and {Chattopadhyay}, Debatri and {Gaebel}, Sebastian M. and {Gittins}, Fabian and {Hirai}, Ryosuke and {Howitt}, George and {Justham}, Stephen and {Khandelwal}, Lokesh and {Kummer}, Floris and {Lau}, Mike Y.~M. and {Mandel}, Ilya and {de Mink}, Selma E. and {Neijssel}, Coenraad and {Riley}, Tim and {van Son}, Lieke and {Stevenson}, Simon and {Vigna-G{\'o}mez}, Alejandro and {Vinciguerra}, Serena and {Wagg}, Tom and {Willcox}, Reinhold},
        title = "{Rapid Stellar and Binary Population Synthesis with COMPAS}",
      journal = {\apjs},
         year = 2022,
        month = feb,
       volume = {258},
       number = {2},
          eid = {34},
        pages = {34},
          doi = {10.3847/1538-4365/ac416c},
       adsurl = {https://ui.adsabs.harvard.edu/abs/2022ApJS..258...34R}
}

@ARTICLE{COMPAS:2025,
       author = {{Team COMPAS: I. Mandel} and {Riley}, Jeff and {Boesky}, Adam and {Brcek}, Adam and {Hirai}, Ryosuke and {Kapil}, Veome and {Lau}, Mike Y.~M. and {Merritt}, JD and {Rodr{\'\i}guez-Segovia}, Nicol{\'a}s and {Romero-Shaw}, Isobel and {Song}, Yuzhe and {Stevenson}, Simon and {Vajpeyi}, Avi and {van Son}, L.~A.~C. and {Vigna-G{\'o}mez}, Alejandro and {Willcox}, Reinhold},
        title = "{Rapid Stellar and Binary Population Synthesis with COMPAS: Methods Paper II}",
      journal = {\apjs},
         year = 2025,
        month = sep,
       volume = {280},
       number = {1},
          eid = {43},
        pages = {43},
          doi = {10.3847/1538-4365/adf8d0},
archivePrefix = {arXiv},
       eprint = {2506.02316},
 primaryClass = {astro-ph.SR},
       adsurl = {https://ui.adsabs.harvard.edu/abs/2025ApJS..280...43T}
}

@ARTICLE{Brown:2001,
       author = {{Brown}, G.~E. and {Heger}, A. and {Langer}, N. and {Lee}, C. -H. and {Wellstein}, S. and {Bethe}, H.~A.},
        title = "{Formation of high mass X-ray black hole binaries}",
      journal = {\na},
         year = 2001,
        month = oct,
       volume = {6},
       number = {7},
        pages = {457-470},
          doi = {10.1016/S1384-1076(01)00077-X},
archivePrefix = {arXiv},
       eprint = {astro-ph/0102379},
 primaryClass = {astro-ph},
       adsurl = {https://ui.adsabs.harvard.edu/abs/2001NewA....6..457B}
}

@ARTICLE{Igoshev:2020,
       author = {{Igoshev}, Andrei P.},
        title = "{The observed velocity distribution of young pulsars - II. Analysis of complete PSR{\ensuremath{\pi}}}",
      journal = {\mnras},
         year = 2020,
        month = may,
       volume = {494},
       number = {3},
        pages = {3663-3674},
          doi = {10.1093/mnras/staa958},
archivePrefix = {arXiv},
       eprint = {2002.01367},
 primaryClass = {astro-ph.HE},
       adsurl = {https://ui.adsabs.harvard.edu/abs/2020MNRAS.494.3663I}
}

@ARTICLE{Kapil:2022,
       author = {{Kapil}, Veome and {Mandel}, Ilya and {Berti}, Emanuele and {M{\"u}ller}, Bernhard},
        title = "{Calibration of neutron star natal kick velocities to isolated pulsar observations}",
      journal = {\mnras},
         year = 2023,
        month = mar,
       volume = {519},
       number = {4},
        pages = {5893-5901},
          doi = {10.1093/mnras/stad019},
archivePrefix = {arXiv},
       eprint = {2209.09252},
 primaryClass = {astro-ph.HE},
       adsurl = {https://ui.adsabs.harvard.edu/abs/2023MNRAS.519.5893K}
}

@ARTICLE{Lynch:2012,
       author = {{Lynch}, Ryan S. and {Freire}, Paulo C.~C. and {Ransom}, Scott M. and {Jacoby}, Bryan A.},
        title = "{The Timing of Nine Globular Cluster Pulsars}",
      journal = {\apj},
         year = 2012,
        month = feb,
       volume = {745},
       number = {2},
          eid = {109},
        pages = {109},
          doi = {10.1088/0004-637X/745/2/109},
archivePrefix = {arXiv},
       eprint = {1112.2612},
 primaryClass = {astro-ph.HE},
       adsurl = {https://ui.adsabs.harvard.edu/abs/2012ApJ...745..109L}
}

@ARTICLE{Burrows:2024,
       author = {{Burrows}, Adam and {Wang}, Tianshu and {Vartanyan}, David},
        title = "{Physical Correlations and Predictions Emerging from Modern Core-Collapse Supernova Theory}",
      journal = {arXiv e-prints},
         year = 2024,
        month = jan,
          eid = {arXiv:2401.06840},
        pages = {arXiv:2401.06840},
          doi = {10.48550/arXiv.2401.06840},
archivePrefix = {arXiv},
       eprint = {2401.06840},
 primaryClass = {astro-ph.HE},
       adsurl = {https://ui.adsabs.harvard.edu/abs/2024arXiv240106840B}
}

@ARTICLE{Schneider:2023,
       author = {{Schneider}, Fabian R.~N. and {Podsiadlowski}, Philipp and {Laplace}, Eva},
        title = "{Bimodal Black Hole Mass Distribution and Chirp Masses of Binary Black Hole Mergers}",
      journal = {\apjl},
         year = 2023,
        month = jun,
       volume = {950},
       number = {2},
          eid = {L9},
        pages = {L9},
          doi = {10.3847/2041-8213/acd77a},
archivePrefix = {arXiv},
       eprint = {2305.02380},
 primaryClass = {astro-ph.HE},
       adsurl = {https://ui.adsabs.harvard.edu/abs/2023ApJ...950L...9S}
}

@ARTICLE{vanSon:2024,
       author = {{van Son}, L.~A.~C. and {Roy}, S.~K. and {Mandel}, I. and {Farr}, W.~M. and {Lam}, A. and {Merritt}, J. and {Broekgaarden}, F.~S. and {Sander}, A.~A.~C. and {Andrews}, J.~J.},
        title = "{Not Just Winds: Why Models Find That Binary Black Hole Formation Is Metallicity-dependent, while Binary Neutron Star Formation Is Not}",
      journal = {\apj},
         year = 2025,
        month = feb,
       volume = {979},
       number = {2},
          eid = {209},
        pages = {209},
          doi = {10.3847/1538-4357/ada14a},
archivePrefix = {arXiv},
       eprint = {2411.02484},
 primaryClass = {astro-ph.HE},
       adsurl = {https://ui.adsabs.harvard.edu/abs/2025ApJ...979..209V}
}

@ARTICLE{Maltsev:2025,
       author = {{Maltsev}, K. and {Schneider}, F.~R.~N. and {Mandel}, I. and {M{\"u}ller}, B. and {Heger}, A. and {R{\"o}pke}, F.~K. and {Laplace}, E.},
        title = "{Explodability criteria for the neutrino-driven supernova mechanism}",
      journal = {arXiv e-prints},
         year = 2025,
        month = mar,
          eid = {arXiv:2503.23856},
        pages = {arXiv:2503.23856},
          doi = {10.48550/arXiv.2503.23856},
archivePrefix = {arXiv},
       eprint = {2503.23856},
 primaryClass = {astro-ph.SR},
       adsurl = {https://ui.adsabs.harvard.edu/abs/2025arXiv250323856M}
}

@ARTICLE{Valli:2025,
       author = {{Valli}, Ruggero and {de Mink}, Selma E. and {Justham}, Stephen and {Callister}, Thomas and {Johnston}, Cole and {Kresse}, Daniel and {Langer}, Norbert and {Rubio}, Amanda C. and {Vigna-G{\'o}mez}, Alejandro and {Wang}, Chen},
        title = "{Evidence of polar and ultralow supernova kicks from the orbits of Be X-ray binaries}",
      journal = {arXiv e-prints},
         year = 2025,
        month = may,
          eid = {arXiv:2505.08857},
        pages = {arXiv:2505.08857},
          doi = {10.48550/arXiv.2505.08857},
archivePrefix = {arXiv},
       eprint = {2505.08857},
 primaryClass = {astro-ph.HE},
       adsurl = {https://ui.adsabs.harvard.edu/abs/2025arXiv250508857V}
}

@ARTICLE{DisbergMandel:2025,
       author = {{Disberg}, Paul and {Mandel}, Ilya},
        title = "{The Kick Velocity Distribution of Isolated Neutron Stars}",
      journal = {\apjl},
         year = 2025,
        month = aug,
       volume = {989},
       number = {1},
          eid = {L8},
        pages = {L8},
          doi = {10.3847/2041-8213/adf286},
archivePrefix = {arXiv},
       eprint = {2505.22102},
 primaryClass = {astro-ph.HE},
       adsurl = {https://ui.adsabs.harvard.edu/abs/2025ApJ...989L...8D}
}

@BOOK{Ivanova:2020,
       author = {{Ivanova}, Natalia and {Justham}, Stephen and {Ricker}, Paul},
        title = "{Common Envelope Evolution}",
         year = 2020,
          doi = {10.1088/2514-3433/abb6f0},
         publisher={IOP Publishing},
       adsurl = {https://ui.adsabs.harvard.edu/abs/2020cee..book.....I}
}

@ARTICLE{Willcox:2025,
       author = {{Willcox}, Reinhold and {Schneider}, Fabian R.~N. and {Laplace}, Eva and {Podsiadlowski}, Philipp and {Maltsev}, Kiril and {Mandel}, Ilya and {Marchant}, Pablo and {Sana}, Hugues and {Li}, Tjonnie G.~F. and {Hertog}, Thomas},
        title = "{Good things always come in 3s: trimodality in the binary black-hole chirp-mass distribution supports bimodal black-hole formation}",
      journal = {arXiv e-prints},
         year = 2025,
        month = oct,
          eid = {arXiv:2510.07573},
        pages = {arXiv:2510.07573},
          doi = {10.48550/arXiv.2510.07573},
archivePrefix = {arXiv},
       eprint = {2510.07573},
 primaryClass = {astro-ph.SR},
       adsurl = {https://ui.adsabs.harvard.edu/abs/2025arXiv251007573W}
}

@ARTICLE{Grichener:2023,
       author = {{Grichener}, Aldana},
        title = "{Mergers of neutron stars and black holes with cores of giant stars: a population synthesis study}",
      journal = {\mnras},
         year = 2023,
        month = jul,
       volume = {523},
       number = {1},
        pages = {221-232},
          doi = {10.1093/mnras/stad1449},
archivePrefix = {arXiv},
       eprint = {2302.06663},
 primaryClass = {astro-ph.HE},
       adsurl = {https://ui.adsabs.harvard.edu/abs/2023MNRAS.523..221G}
}

@ARTICLE{Manchester:2005,
       author = {{Manchester}, R.~N. and {Hobbs}, G.~B. and {Teoh}, A. and {Hobbs}, M.},
        title = "{The Australia Telescope National Facility Pulsar Catalogue}",
      journal = {\aj},
         year = 2005,
        month = apr,
       volume = {129},
       number = {4},
        pages = {1993-2006},
          doi = {10.1086/428488},
archivePrefix = {arXiv},
       eprint = {astro-ph/0412641},
 primaryClass = {astro-ph},
       adsurl = {https://ui.adsabs.harvard.edu/abs/2005AJ....129.1993M}
}

@ARTICLE{Ding:2024,
       author = {{Ding}, Hao and {Deller}, Adam T. and {Swiggum}, Joseph K. and {Lynch}, Ryan S. and {Chatterjee}, Shami and {Tauris}, Thomas M.},
        title = "{VLBA Astrometry of the Galactic Double Neutron Stars PSR J0509+3801 and PSR J1930{\textendash}1852: A Preliminary Transverse Velocity Distribution of Double Neutron Stars and its Implications}",
      journal = {\apj},
         year = 2024,
        month = jul,
       volume = {970},
       number = {1},
          eid = {90},
        pages = {90},
          doi = {10.3847/1538-4357/ad4883},
archivePrefix = {arXiv},
       eprint = {2405.03914},
 primaryClass = {astro-ph.HE},
       adsurl = {https://ui.adsabs.harvard.edu/abs/2024ApJ...970...90D}
}

@ARTICLE{Lyne:1996,
       author = {{Lyne}, A.~G. and {Pritchard}, R.~S. and {Graham-Smith}, F. and {Camilo}, F.},
        title = "{Very low braking index for the Vela pulsar}",
      journal = {\nat},
         year = 1996,
        month = jun,
       volume = {381},
       number = {6582},
        pages = {497-498},
          doi = {10.1038/381497a0},
       adsurl = {https://ui.adsabs.harvard.edu/abs/1996Natur.381..497L}
}

@ARTICLE{Disberg:2024a,
       author = {{Disberg}, P. and {Gaspari}, N. and {Levan}, A.~J.},
        title = "{Deceleration of kicked objects due to the Galactic potential}",
      journal = {\aap},
         year = 2024,
        month = jul,
       volume = {687},
          eid = {A272},
        pages = {A272},
          doi = {10.1051/0004-6361/202449996},
archivePrefix = {arXiv},
       eprint = {2405.06436},
 primaryClass = {astro-ph.GA},
       adsurl = {https://ui.adsabs.harvard.edu/abs/2024A&A...687A.272D}
}

@ARTICLE{Disberg:2024b,
       author = {{Disberg}, P. and {Gaspari}, N. and {Levan}, A.~J.},
        title = "{Kinematic constraints on the ages and kick velocities of Galactic neutron star binaries}",
      journal = {\aap},
         year = 2024,
        month = sep,
       volume = {689},
          eid = {A348},
        pages = {A348},
          doi = {10.1051/0004-6361/202450790},
       adsurl = {https://ui.adsabs.harvard.edu/abs/2024A&A...689A.348D}
}

@ARTICLE{Disberg:2025,
       author = {{Disberg}, P. and {Gaspari}, N. and {Levan}, A.~J.},
        title = "{A kinematically constrained kick distribution for isolated neutron stars}",
      journal = {\aap},
         year = 2025,
        month = aug,
       volume = {700},
          eid = {A75},
        pages = {A75},
          doi = {10.1051/0004-6361/202554349},
archivePrefix = {arXiv},
       eprint = {2503.01429},
 primaryClass = {astro-ph.HE},
       adsurl = {https://ui.adsabs.harvard.edu/abs/2025A&A...700A..75D}
}

@ARTICLE{Igoshev:2019,
       author = {{Igoshev}, Andrei P.},
        title = "{Ages of radio pulsar: long-term magnetic field evolution}",
      journal = {\mnras},
         year = 2019,
        month = jan,
       volume = {482},
       number = {3},
        pages = {3415-3425},
          doi = {10.1093/mnras/sty2945},
archivePrefix = {arXiv},
       eprint = {1810.12922},
 primaryClass = {astro-ph.HE},
       adsurl = {https://ui.adsabs.harvard.edu/abs/2019MNRAS.482.3415I}
}

@ARTICLE{Maoz:2024,
       author = {{Maoz}, Dan and {Nakar}, Ehud},
        title = "{The Neutron Star Merger Delay-time Distribution, R-process ``Knees,'' and the Metal Budget of the Galaxy}",
      journal = {\apj},
         year = 2025,
        month = apr,
       volume = {982},
       number = {2},
          eid = {179},
        pages = {179},
          doi = {10.3847/1538-4357/ada3bd},
archivePrefix = {arXiv},
       eprint = {2406.08630},
 primaryClass = {astro-ph.HE},
       adsurl = {https://ui.adsabs.harvard.edu/abs/2025ApJ...982..179M}
}

@ARTICLE{Barr:2024,
       author = {{Barr}, Ewan D. and {Dutta}, Arunima and {Freire}, Paulo C.~C. and {Cadelano}, Mario and {Gautam}, Tasha and {Kramer}, Michael and {Pallanca}, Cristina and {Ransom}, Scott M. and {Ridolfi}, Alessandro and {Stappers}, Benjamin W. and {Tauris}, Thomas M. and {Venkatraman Krishnan}, Vivek and {Wex}, Norbert and {Bailes}, Matthew and {Behrend}, Jan and {Buchner}, Sarah and {Burgay}, Marta and {Chen}, Weiwei and {Champion}, David J. and {Chen}, C.-H. Rosie and {Corongiu}, Alessandro and {Geyer}, Marisa and {Men}, Y.~P. and {Padmanabh}, Prajwal Voraganti and {Possenti}, Andrea},
        title = "{A pulsar in a binary with a compact object in the mass gap between neutron stars and black holes}",
      journal = {Science},
         year = 2024,
        month = jan,
       volume = {383},
       number = {6680},
        pages = {275-279},
          doi = {10.1126/science.adg3005},
archivePrefix = {arXiv},
       eprint = {2401.09872},
 primaryClass = {astro-ph.HE},
       adsurl = {https://ui.adsabs.harvard.edu/abs/2024Sci...383..275B}
}

@ARTICLE{Keith:2009,
       author = {{Keith}, M.~J. and {Kramer}, M. and {Lyne}, A.~G. and {Eatough}, R.~P. and {Stairs}, I.~H. and {Possenti}, A. and {Camilo}, F. and {Manchester}, R.~N.},
        title = "{PSR J1753-2240: a mildly recycled pulsar in an eccentric binary system}",
      journal = {\mnras},
         year = 2009,
        month = feb,
       volume = {393},
       number = {2},
        pages = {623-627},
          doi = {10.1111/j.1365-2966.2008.14234.x},
archivePrefix = {arXiv},
       eprint = {0811.2027},
 primaryClass = {astro-ph},
       adsurl = {https://ui.adsabs.harvard.edu/abs/2009MNRAS.393..623K}
}

@ARTICLE{Ng:2018b,
       author = {{Ng}, C. and {Kruckow}, M.~U. and {Tauris}, T.~M. and {Lyne}, A.~G. and {Freire}, P.~C.~C. and {Ridolfi}, A. and {Caiazzo}, I. and {Heyl}, J. and {Kramer}, M. and {Cameron}, A.~D. and {Champion}, D.~J. and {Stappers}, B.},
        title = "{PSR J1755-2550: a young radio pulsar with a massive, compact companion}",
      journal = {\mnras},
         year = 2018,
        month = jun,
       volume = {476},
       number = {4},
        pages = {4315-4326},
          doi = {10.1093/mnras/sty482},
archivePrefix = {arXiv},
       eprint = {1802.08248},
 primaryClass = {astro-ph.HE},
       adsurl = {https://ui.adsabs.harvard.edu/abs/2018MNRAS.476.4315N}
}

@ARTICLE{Van_Leeuwen:2015,
       author = {{van Leeuwen}, J. and {Kasian}, L. and {Stairs}, I.~H. and {Lorimer}, D.~R. and {Camilo}, F. and {Chatterjee}, S. and {Cognard}, I. and {Desvignes}, G. and {Freire}, P.~C.~C. and {Janssen}, G.~H. and {Kramer}, M. and {Lyne}, A.~G. and {Nice}, D.~J. and {Ransom}, S.~M. and {Stappers}, B.~W. and {Weisberg}, J.~M.},
        title = "{The Binary Companion of Young, Relativistic Pulsar J1906+0746}",
      journal = {\apj},
         year = 2015,
        month = jan,
       volume = {798},
       number = {2},
          eid = {118},
        pages = {118},
          doi = {10.1088/0004-637X/798/2/118},
archivePrefix = {arXiv},
       eprint = {1411.1518},
 primaryClass = {astro-ph.SR},
       adsurl = {https://ui.adsabs.harvard.edu/abs/2015ApJ...798..118V}
}

@ARTICLE{Phinney:1991,
       author = {{Phinney}, E.~S. and {Sigurdsson}, Steinn},
        title = "{Ejection of pulsars and binaries to the outskirts of globular clusters}",
      journal = {\nat},
         year = 1991,
        month = jan,
       volume = {349},
       number = {6306},
        pages = {220-223},
          doi = {10.1038/349220a0},
       adsurl = {https://ui.adsabs.harvard.edu/abs/1991Natur.349..220P}
}

@ARTICLE{Chattaraj:2025,
       author = {{Chattaraj}, Abhishek and {Andrews}, Jeff J. and {Bavera}, Simone S. and {Briel}, Max and {Chattopadhyay}, Debatri and {Fragos}, Tassos and {Gossage}, Seth and {Kalogera}, Vicky and {Kovlakas}, Konstantinos and {Kruckow}, Matthias U. and {Liotine}, Camille and {Rocha}, Kyle A. and {Srivastava}, Philipp M. and {Sun}, Meng and {Teng}, Elizabeth and {Xing}, Zepei and {Zapartas}, Emmanouil},
        title = "{Forming Double Neutron Stars Using Detailed Binary Evolution Models with POSYDON: Comparison to the Galactic Systems}",
      journal = {\apj},
         year = 2026,
        month = jan,
       volume = {997},
       number = {1},
          eid = {52},
        pages = {52},
          doi = {10.3847/1538-4357/ae1f93},
archivePrefix = {arXiv},
       eprint = {2508.00186},
 primaryClass = {astro-ph.SR},
       adsurl = {https://ui.adsabs.harvard.edu/abs/2026ApJ...997...52C}
}

@ARTICLE{BoccioliFragione:2024,
       author = {{Boccioli}, Luca and {Fragione}, Giacomo},
        title = "{Remnant masses from 1D+ core-collapse supernovae simulations: Bimodal neutron star mass distribution and black holes in the low-mass gap}",
      journal = {\prd},
         year = 2024,
        month = jul,
       volume = {110},
       number = {2},
          eid = {023007},
        pages = {023007},
          doi = {10.1103/PhysRevD.110.023007},
archivePrefix = {arXiv},
       eprint = {2404.05927},
 primaryClass = {astro-ph.HE},
       adsurl = {https://ui.adsabs.harvard.edu/abs/2024PhRvD.110b3007B}
}

@ARTICLE{Boccioli:2025,
       author = {{Boccioli}, Luca and {Vartanyan}, David and {O'Connor}, Evan P. and {Kasen}, Daniel},
        title = "{Neutrino heating in 1D, 2D, and 3D core-collapse supernovae: characterizing the explosion of high-compactness stars}",
      journal = {\mnras},
         year = 2025,
        month = jul,
       volume = {540},
       number = {4},
        pages = {3885-3905},
          doi = {10.1093/mnras/staf963},
archivePrefix = {arXiv},
       eprint = {2501.06784},
 primaryClass = {astro-ph.HE},
       adsurl = {https://ui.adsabs.harvard.edu/abs/2025MNRAS.540.3885B}
}

@ARTICLE{Nair:2025,
       author = {{Nair}, Ashwathi and {Stevenson}, Simon},
        title = "{Formation of heavy double neutron stars {\textendash} I. Eddington-limited accretion for a 1.4 M$_{{\ensuremath{\odot}}}$ neutron star at solar metallicity}",
      journal = {\mnras},
         year = 2025,
        month = oct,
       volume = {543},
       number = {1},
        pages = {233-246},
          doi = {10.1093/mnras/staf1397},
archivePrefix = {arXiv},
       eprint = {2508.15624},
 primaryClass = {astro-ph.HE},
       adsurl = {https://ui.adsabs.harvard.edu/abs/2025MNRAS.543..233N}
}

@ARTICLE{Abbott:2020,
       author = {{Abbott}, B.~P. and {Abbott}, R. and {Abbott}, T.~D. and {Abraham}, S. and {Acernese}, F. and {Ackley}, K. and {Adams}, C. and {Adhikari}, R.~X. and {Adya}, V.~B. and {Affeldt}, C. and {Agathos}, M. and {Agatsuma}, K. and {Aggarwal}, N. and {Aguiar}, O.~D. and {Aiello}, L. and {Ain}, A. and {Ajith}, P. and {Allen}, G. and {Allocca}, A. and {Aloy}, M.~A. and {Altin}, P.~A. and {Amato}, A. and {Anand}, S. and {Ananyeva}, A. and {Anderson}, S.~B. and {Anderson}, W.~G. and {Angelova}, S.~V. and {Antier}, S. and {Appert}, S. and {Arai}, K. and {Araya}, M.~C. and {Areeda}, J.~S. and {Ar{\`e}ne}, M. and {Arnaud}, N. and {Aronson}, S.~M. and {Arun}, K.~G. and {Ascenzi}, S. and {Ashton}, G. and {Aston}, S.~M. and {Astone}, P. and {Aubin}, F. and {Aufmuth}, P. and {AultONeal}, K. and {Austin}, C. and {Avendano}, V. and {Avila-Alvarez}, A. and {Babak}, S. and {Bacon}, P. and {Badaracco}, F. and {Bader}, M.~K.~M. and {Bae}, S. and {Baird}, J. and {Baker}, P.~T. and {Baldaccini}, F. and {Ballardin}, G. and {Ballmer}, S.~W. and {Bals}, A. and {Banagiri}, S. and {Barayoga}, J.~C. and {Barbieri}, C. and {Barclay}, S.~E. and {Barish}, B.~C. and {Barker}, D. and {Barkett}, K. and {Barnum}, S. and {Barone}, F. and {Barr}, B. and {Barsotti}, L. and {Barsuglia}, M. and {Barta}, D. and {Bartlett}, J. and {Bartos}, I. and {Bassiri}, R. and {Basti}, A. and {Bawaj}, M. and {Bayley}, J.~C. and {Baylor}, A.~C. and {Bazzan}, M. and {B{\'e}csy}, B. and {Bejger}, M. and {Belahcene}, I. and {Bell}, A.~S. and {Beniwal}, D. and {Benjamin}, M.~G. and {Berger}, B.~K. and {Bergmann}, G. and {Bernuzzi}, S. and {Berry}, C.~P.~L. and {Bersanetti}, D. and {Bertolini}, A. and {Betzwieser}, J. and {Bhandare}, R. and {Bidler}, J. and {Biggs}, E. and {Bilenko}, I.~A. and {Bilgili}, S.~A. and {Billingsley}, G. and {Birney}, R. and {Birnholtz}, O. and {Biscans}, S. and {Bischi}, M. and {Biscoveanu}, S. and {Bisht}, A. and {Bitossi}, M. and {Bizouard}, M.~A. and {Blackburn}, J.~K. and {Blackman}, J. and {Blair}, C.~D. and {Blair}, D.~G. and {Blair}, R.~M. and {Bloemen}, S. and {Bobba}, F. and {Bode}, N. and {Boer}, M. and {Boetzel}, Y. and {Bogaert}, G. and {Bondu}, F. and {Bonnand}, R. and {Booker}, P. and {Boom}, B.~A. and {Bork}, R. and {Boschi}, V. and {Bose}, S. and {Bossilkov}, V. and {Bosveld}, J. and {Bouffanais}, Y. and {Bozzi}, A. and {Bradaschia}, C. and {Brady}, P.~R. and {Bramley}, A. and {Branchesi}, M. and {Brau}, J.~E. and {Breschi}, M. and {Briant}, T. and {Briggs}, J.~H. and {Brighenti}, F. and {Brillet}, A. and {Brinkmann}, M. and {Brockill}, P. and {Brooks}, A.~F. and {Brooks}, J. and {Brown}, D.~D. and {Brunett}, S. and {Buikema}, A. and {Bulik}, T. and {Bulten}, H.~J. and {Buonanno}, A. and {Buskulic}, D. and {Buy}, C. and {Byer}, R.~L. and {Cabero}, M. and {Cadonati}, L. and {Cagnoli}, G. and {Cahillane}, C. and {Calder{\'o}n Bustillo}, J. and {Callister}, T.~A. and {Calloni}, E. and {Camp}, J.~B. and {Campbell}, W.~A. and {Canepa}, M. and {Cannon}, K.~C. and {Cao}, H. and {Cao}, J. and {Carapella}, G. and {Carbognani}, F. and {Caride}, S. and {Carney}, M.~F. and {Carullo}, G. and {Casanueva Diaz}, J. and {Casentini}, C. and {Caudill}, S. and {Cavagli{\`a}}, M. and {Cavalier}, F. and {Cavalieri}, R. and {Cella}, G. and {Cerd{\'a}-Dur{\'a}n}, P. and {Cesarini}, E. and {Chaibi}, O. and {Chakravarti}, K. and {Chamberlin}, S.~J. and {Chan}, M. and {Chao}, S. and {Charlton}, P. and {Chase}, E.~A. and {Chassande-Mottin}, E. and {Chatterjee}, D. and {Chaturvedi}, M. and {Chatziioannou}, K. and {Cheeseboro}, B.~D. and {Chen}, H.~Y. and {Chen}, X. and {Chen}, Y. and {Cheng}, H.-P. and {Cheong}, C.~K. and {Chia}, H.~Y. and {Chiadini}, F. and {Chincarini}, A. and {Chiummo}, A. and {Cho}, G. and {Cho}, H.~S.},
        title = "{GW190425: Observation of a Compact Binary Coalescence with Total Mass {\ensuremath{\sim}} 3.4 M$_{{\ensuremath{\odot}}}$}",
      journal = {\apjl},
         year = 2020,
        month = mar,
       volume = {892},
       number = {1},
          eid = {L3},
        pages = {L3},
          doi = {10.3847/2041-8213/ab75f5},
archivePrefix = {arXiv},
       eprint = {2001.01761},
 primaryClass = {astro-ph.HE},
       adsurl = {https://ui.adsabs.harvard.edu/abs/2020ApJ...892L...3A}
}

@ARTICLE{Grichener:2025,
       author = {{Grichener}, Aldana},
        title = "{Mergers of compact objects with cores of massive stars: evolutionary pathways, r-process nucleosynthesis and multi-messenger signatures}",
      journal = {\apss},
         year = 2025,
        month = feb,
       volume = {370},
       number = {2},
          eid = {11},
        pages = {11},
          doi = {10.1007/s10509-025-04402-1},
archivePrefix = {arXiv},
       eprint = {2410.18813},
 primaryClass = {astro-ph.HE},
       adsurl = {https://ui.adsabs.harvard.edu/abs/2025Ap&SS.370...11G}
}

@ARTICLE{Bernadich:2023,
       author = {{Colom i Bernadich}, M. and {Balakrishnan}, V. and {Barr}, E. and {Berezina}, M. and {Burgay}, M. and {Buchner}, S. and {Champion}, D.~J. and {Chen}, W. and {Desvignes}, G. and {Freire}, P.~C.~C. and {Grunthal}, K. and {Kramer}, M. and {Men}, Y. and {Padmanabh}, P.~V. and {Parthasarathy}, A. and {Pillay}, D. and {Rammala}, I. and {Sengupta}, S. and {Venkatraman Krishnan}, V.},
        title = "{The MPIfR-MeerKAT Galactic Plane Survey. II. The eccentric double neutron star system PSR J1208{\ensuremath{-}}5936 and a neutron star merger rate update}",
      journal = {\aap},
         year = 2023,
        month = oct,
       volume = {678},
          eid = {A187},
        pages = {A187},
          doi = {10.1051/0004-6361/202346953},
archivePrefix = {arXiv},
       eprint = {2308.16802},
 primaryClass = {astro-ph.HE},
       adsurl = {https://ui.adsabs.harvard.edu/abs/2023A&A...678A.187C}
}

@ARTICLE{Moriya:2017,
       author = {{Moriya}, Takashi J. and {Mazzali}, Paolo A. and {Tominaga}, Nozomu and {Hachinger}, Stephan and {Blinnikov}, Sergei I. and {Tauris}, Thomas M. and {Takahashi}, Koh and {Tanaka}, Masaomi and {Langer}, Norbert and {Podsiadlowski}, Philipp},
        title = "{Light-curve and spectral properties of ultrastripped core-collapse supernovae leading to binary neutron stars}",
      journal = {\mnras},
         year = 2017,
        month = apr,
       volume = {466},
       number = {2},
        pages = {2085-2098},
          doi = {10.1093/mnras/stw3225},
archivePrefix = {arXiv},
       eprint = {1612.02882},
 primaryClass = {astro-ph.HE},
       adsurl = {https://ui.adsabs.harvard.edu/abs/2017MNRAS.466.2085M}
}

@ARTICLE{De:2018,
       author = {{De}, K. and {Kasliwal}, M.~M. and {Ofek}, E.~O. and {Moriya}, T.~J. and {Burke}, J. and {Cao}, Y. and {Cenko}, S.~B. and {Doran}, G.~B. and {Duggan}, G.~E. and {Fender}, R.~P. and {Fransson}, C. and {Gal-Yam}, A. and {Horesh}, A. and {Kulkarni}, S.~R. and {Laher}, R.~R. and {Lunnan}, R. and {Manulis}, I. and {Masci}, F. and {Mazzali}, P.~A. and {Nugent}, P.~E. and {Perley}, D.~A. and {Petrushevska}, T. and {Piro}, A.~L. and {Rumsey}, C. and {Sollerman}, J. and {Sullivan}, M. and {Taddia}, F.},
        title = "{A hot and fast ultra-stripped supernova that likely formed a compact neutron star binary}",
      journal = {Science},
         year = 2018,
        month = oct,
       volume = {362},
       number = {6411},
        pages = {201-206},
          doi = {10.1126/science.aas8693},
archivePrefix = {arXiv},
       eprint = {1810.05181},
 primaryClass = {astro-ph.HE},
       adsurl = {https://ui.adsabs.harvard.edu/abs/2018Sci...362..201D}
}

@ARTICLE{Laplace:2025,
       author = {{Laplace}, E. and {Schneider}, F.~R.~N. and {Podsiadlowski}, Ph.},
        title = "{It's written in the massive stars: The role of stellar physics in the formation of black holes}",
      journal = {\aap},
         year = 2025,
        month = mar,
       volume = {695},
          eid = {A71},
        pages = {A71},
          doi = {10.1051/0004-6361/202451077},
archivePrefix = {arXiv},
       eprint = {2409.02058},
 primaryClass = {astro-ph.SR},
       adsurl = {https://ui.adsabs.harvard.edu/abs/2025A&A...695A..71L}
}

@ARTICLE{Wong:2010,
       author = {{Wong}, Tsing-Wai and {Willems}, Bart and {Kalogera}, Vassiliki},
        title = "{Constraints on Natal Kicks in Galactic Double Neutron Star Systems}",
      journal = {\apj},
         year = 2010,
        month = oct,
       volume = {721},
       number = {2},
        pages = {1689-1701},
          doi = {10.1088/0004-637X/721/2/1689},
archivePrefix = {arXiv},
       eprint = {1008.2397},
 primaryClass = {astro-ph.HE},
       adsurl = {https://ui.adsabs.harvard.edu/abs/2010ApJ...721.1689W}
}

@ARTICLE{Jacoby:2006,
       author = {{Jacoby}, B.~A. and {Cameron}, P.~B. and {Jenet}, F.~A. and {Anderson}, S.~B. and {Murty}, R.~N. and {Kulkarni}, S.~R.},
        title = "{Measurement of Orbital Decay in the Double Neutron Star Binary PSR B2127+11C}",
      journal = {\apjl},
         year = 2006,
        month = jun,
       volume = {644},
       number = {2},
        pages = {L113-L116},
          doi = {10.1086/505742},
archivePrefix = {arXiv},
       eprint = {astro-ph/0605375},
 primaryClass = {astro-ph},
       adsurl = {https://ui.adsabs.harvard.edu/abs/2006ApJ...644L.113J}
}

@ARTICLE{Disberg:2026,
       author = {{Disberg}, Paul and {Bahramian}, Arash and {Mandel}, Ilya},
        title = "{Reconciling the Systemic Kicks of Observed Millisecond Pulsars, Spider Pulsars, and Low-mass X-Ray Binaries}",
      journal = {\apjl},
         year = 2026,
        month = apr,
       volume = {1000},
       number = {2},
          eid = {L56},
        pages = {L56},
          doi = {10.3847/2041-8213/ae52f1},
archivePrefix = {arXiv},
       eprint = {2601.12275},
 primaryClass = {astro-ph.HE},
       adsurl = {https://ui.adsabs.harvard.edu/abs/2026ApJ..1000L..56D}
}

@ARTICLE{Ferdman:2013,
       author = {{Ferdman}, R.~D. and {Stairs}, I.~H. and {Kramer}, M. and {Breton}, R.~P. and {McLaughlin}, M.~A. and {Freire}, P.~C.~C. and {Possenti}, A. and {Stappers}, B.~W. and {Kaspi}, V.~M. and {Manchester}, R.~N. and {Lyne}, A.~G.},
        title = "{The Double Pulsar: Evidence for Neutron Star Formation without an Iron Core-collapse Supernova}",
      journal = {\apj},
         year = 2013,
        month = apr,
       volume = {767},
       number = {1},
          eid = {85},
        pages = {85},
          doi = {10.1088/0004-637X/767/1/85},
archivePrefix = {arXiv},
       eprint = {1302.2914},
 primaryClass = {astro-ph.SR},
       adsurl = {https://ui.adsabs.harvard.edu/abs/2013ApJ...767...85F}
}

@ARTICLE{Kaaret:2017,
       author = {{Kaaret}, Philip and {Feng}, Hua and {Roberts}, Timothy P.},
        title = "{Ultraluminous X-Ray Sources}",
      journal = {\araa},
         year = 2017,
        month = aug,
       volume = {55},
       number = {1},
        pages = {303-341},
          doi = {10.1146/annurev-astro-091916-055259},
archivePrefix = {arXiv},
       eprint = {1703.10728},
 primaryClass = {astro-ph.HE},
       adsurl = {https://ui.adsabs.harvard.edu/abs/2017ARA&A..55..303K}
}

@ARTICLE{Israel:2017,
       author = {{Israel}, Gian Luca and {Belfiore}, Andrea and {Stella}, Luigi and {Esposito}, Paolo and {Casella}, Piergiorgio and {De Luca}, Andrea and {Marelli}, Martino and {Papitto}, Alessandro and {Perri}, Matteo and {Puccetti}, Simonetta and {Castillo}, Guillermo A. Rodr{\'\i}guez and {Salvetti}, David and {Tiengo}, Andrea and {Zampieri}, Luca and {D'Agostino}, Daniele and {Greiner}, Jochen and {Haberl}, Frank and {Novara}, Giovanni and {Salvaterra}, Ruben and {Turolla}, Roberto and {Watson}, Mike and {Wilms}, Joern and {Wolter}, Anna},
        title = "{An accreting pulsar with extreme properties drives an ultraluminous x-ray source in NGC 5907}",
      journal = {Science},
         year = 2017,
        month = feb,
       volume = {355},
       number = {6327},
        pages = {817-819},
          doi = {10.1126/science.aai8635},
archivePrefix = {arXiv},
       eprint = {1609.07375},
 primaryClass = {astro-ph.HE},
       adsurl = {https://ui.adsabs.harvard.edu/abs/2017Sci...355..817I}
}

@ARTICLE{Tauris:2012,
       author = {{Tauris}, T.~M. and {Langer}, N. and {Kramer}, M.},
        title = "{Formation of millisecond pulsars with CO white dwarf companions - II. Accretion, spin-up, true ages and comparison to MSPs with He white dwarf companions}",
      journal = {\mnras},
         year = 2012,
        month = sep,
       volume = {425},
       number = {3},
        pages = {1601-1627},
          doi = {10.1111/j.1365-2966.2012.21446.x},
archivePrefix = {arXiv},
       eprint = {1206.1862},
 primaryClass = {astro-ph.SR},
       adsurl = {https://ui.adsabs.harvard.edu/abs/2012MNRAS.425.1601T}
}

@article{massey1951kolmogorov,
  title={The Kolmogorov-Smirnov test for goodness of fit},
  author={Massey Jr, Frank J},
  journal={Journal of the American statistical Association},
  volume={46},
  number={253},
  pages={68--78},
  year={1951},
  publisher={Taylor \& Francis}
}

@ARTICLE{Braun:2019,
       author = {{Braun}, Robert and {Bonaldi}, Anna and {Bourke}, Tyler and {Keane}, Evan and {Wagg}, Jeff},
        title = "{Anticipated Performance of the Square Kilometre Array -- Phase 1 (SKA1)}",
      journal = {arXiv e-prints},
         year = 2019,
        month = dec,
          eid = {arXiv:1912.12699},
        pages = {arXiv:1912.12699},
          doi = {10.48550/arXiv.1912.12699},
archivePrefix = {arXiv},
       eprint = {1912.12699},
 primaryClass = {astro-ph.IM},
       adsurl = {https://ui.adsabs.harvard.edu/abs/2019arXiv191212699B}
}

@ARTICLE{Yang:2026,
       author = {{Yang}, Z.~L. and {Han}, J.~L. and {Wang}, P.~F. and {Wang}, C. and {Cai}, N.~N. and {Jing}, W.~C. and {Su}, W.~Q. and {Wang}, T. and {Xu}, J. and {Yan}, Yi and {Zhou}, D.~J.},
        title = "{Mass measurements of the double neutron star system PSR J0641+0448}",
      journal = {\apjl},
        year = {2026},
        month = {mar},
        publisher = {The American Astronomical Society},
        volume = {1000},
        number = {1},
        pages = {L1},
          doi = {10.3847/2041-8213/ae4801},
       adsurl = {https://ui.adsabs.harvard.edu/abs/2026arXiv260310788Y}
}

@ARTICLE{Woosley:2019,
       author = {{Woosley}, S.~E.},
        title = "{The Evolution of Massive Helium Stars, Including Mass Loss}",
      journal = {\apj},
         year = 2019,
        month = jun,
       volume = {878},
       number = {1},
          eid = {49},
        pages = {49},
          doi = {10.3847/1538-4357/ab1b41},
archivePrefix = {arXiv},
       eprint = {1901.00215},
 primaryClass = {astro-ph.SR},
       adsurl = {https://ui.adsabs.harvard.edu/abs/2019ApJ...878...49W}
}

@ARTICLE{Shishkin:2023,
       author = {{Shishkin}, Dmitry and {Soker}, Noam},
        title = "{Remnant masses of core collapse supernovae in the jittering jets explosion mechanism}",
      journal = {\mnras},
         year = 2022,
        month = jul,
       volume = {513},
       number = {3},
        pages = {4224-4231},
          doi = {10.1093/mnras/stac1075},
archivePrefix = {arXiv},
       eprint = {2112.10697},
 primaryClass = {astro-ph.HE},
       adsurl = {https://ui.adsabs.harvard.edu/abs/2022MNRAS.513.4224S}
}

@ARTICLE{Zhao:2024,
       author = {{Zhao}, D. and {Wang}, N. and {Yuan}, J.~P. and {Li}, D. and {Wang}, P. and {Xue}, M.~Y. and {Zhu}, W.~W. and {Miao}, C.~C. and {Yan}, W.~M. and {Wang}, J.~B. and {Yao}, J.~M. and {Wu}, Q.~D. and {Wang}, S.~Q. and {Sun}, S.~N. and {Kou}, F.~F. and {Chen}, Y.~T. and {Dang}, S.~J. and {Feng}, Y. and {Liu}, Z.~J. and {Miao}, X.~L. and {Meng}, L.~Q. and {Yuan}, M. and {Niu}, C.~H. and {Niu}, J.~R. and {Qian}, L. and {Wang}, S. and {Xie}, X.~Y. and {Xiao}, Y.~F. and {Yue}, Y.~L. and {You}, S.~P. and {Yu}, X.~H. and {Zhao}, R.~S. and {Yuen}, R. and {Zhou}, X. and {Zhang}, L. and {Wang}, Y.~B. and {Wu}, J.~F. and {Gan}, Z.~Y. and {Sun}, Z.~Y. and {Wang}, C.~J.},
        title = "{A Relativistic Double Neutron Star Binary PSR J1846-0513}",
      journal = {\apjl},
         year = 2024,
        month = mar,
       volume = {964},
       number = {1},
          eid = {L7},
        pages = {L7},
          doi = {10.3847/2041-8213/ad2fb3},
       adsurl = {https://ui.adsabs.harvard.edu/abs/2024ApJ...964L...7Z}
}

@ARTICLE{Wong2010,
       author = {{Wong}, Tsing-Wai and {Willems}, Bart and {Kalogera}, Vassiliki},
        title = "{Constraints on Natal Kicks in Galactic Double Neutron Star Systems}",
      journal = {\apj},
         year = 2010,
        month = oct,
       volume = {721},
       number = {2},
        pages = {1689-1701},
          doi = {10.1088/0004-637X/721/2/1689},
archivePrefix = {arXiv},
       eprint = {1008.2397},
 primaryClass = {astro-ph.HE},
       adsurl = {https://ui.adsabs.harvard.edu/abs/2010ApJ...721.1689W}
}

@ARTICLE{Thorsett2005,
       author = {{Thorsett}, S.~E. and {Dewey}, R.~J. and {Stairs}, I.~H.},
        title = "{Studies of the Relativistic Binary Pulsar PSR B1534+12. II. Origin and Evolution}",
      journal = {\apj},
         year = 2005,
        month = feb,
       volume = {619},
       number = {2},
        pages = {1036-1045},
          doi = {10.1086/426668},
archivePrefix = {arXiv},
       eprint = {astro-ph/0408458},
 primaryClass = {astro-ph},
       adsurl = {https://ui.adsabs.harvard.edu/abs/2005ApJ...619.1036T}
}

@ARTICLE{DisbergNelemans:2023,
       author = {{Disberg}, P. and {Nelemans}, G.},
        title = "{Failed supernovae as a natural explanation for the binary black hole mass distribution}",
      journal = {\aap},
         year = 2023,
        month = aug,
       volume = {676},
          eid = {A31},
        pages = {A31},
          doi = {10.1051/0004-6361/202245693},
archivePrefix = {arXiv},
       eprint = {2306.14332},
 primaryClass = {astro-ph.HE},
       adsurl = {https://ui.adsabs.harvard.edu/abs/2023A&A...676A..31D}
}

@ARTICLE{Timmes1996,
       author = {{Timmes}, F.~X. and {Woosley}, S.~E. and {Weaver}, Thomas A.},
        title = "{The Neutron Star and Black Hole Initial Mass Function}",
      journal = {\apj},
         year = 1996,
        month = feb,
       volume = {457},
        pages = {834},
          doi = {10.1086/176778},
archivePrefix = {arXiv},
       eprint = {astro-ph/9510136},
 primaryClass = {astro-ph},
       adsurl = {https://ui.adsabs.harvard.edu/abs/1996ApJ...457..834T}
}

@ARTICLE{Sukhbold:2020,
       author = {{Sukhbold}, Tuguldur and {Adams}, Scott},
        title = "{Missing red supergiants and carbon burning}",
      journal = {\mnras},
         year = 2020,
        month = feb,
       volume = {492},
       number = {2},
        pages = {2578-2587},
          doi = {10.1093/mnras/staa059},
archivePrefix = {arXiv},
       eprint = {1905.00474},
 primaryClass = {astro-ph.HE},
       adsurl = {https://ui.adsabs.harvard.edu/abs/2020MNRAS.492.2578S}
}

@ARTICLE{Summa:2016,
       author = {{Summa}, Alexander and {Hanke}, Florian and {Janka}, Hans-Thomas and {Melson}, Tobias and {Marek}, Andreas and {M{\"u}ller}, Bernhard},
        title = "{Progenitor-dependent Explosion Dynamics in Self-consistent, Axisymmetric Simulations of Neutrino-driven Core-collapse Supernovae}",
      journal = {\apj},
         year = 2016,
        month = jul,
       volume = {825},
       number = {1},
          eid = {6},
        pages = {6},
          doi = {10.3847/0004-637X/825/1/6},
archivePrefix = {arXiv},
       eprint = {1511.07871},
 primaryClass = {astro-ph.SR},
       adsurl = {https://ui.adsabs.harvard.edu/abs/2016ApJ...825....6S}
}

@ARTICLE{YamamotoYamada2016,
       author = {{Yamamoto}, Yu and {Yamada}, Shoichi},
        title = "{Systematic Studies of Shock Revival and the Subsequent Evolutions in Core-collapse Supernovae with Parametric Progenitor Models}",
      journal = {\apj},
         year = 2016,
        month = feb,
       volume = {818},
       number = {2},
          eid = {165},
        pages = {165},
          doi = {10.3847/0004-637X/818/2/165},
       adsurl = {https://ui.adsabs.harvard.edu/abs/2016ApJ...818..165Y}
}

@ARTICLE{delPeloso2005,
       author = {{del Peloso}, E.~F. and {da Silva}, L. and {Porto de Mello}, G.~F. and {Arany-Prado}, L.~I.},
        title = "{The age of the Galactic thin disk from Th/Eu nucleocosmochronology. III. Extended sample}",
      journal = {\aap},
         year = 2005,
        month = sep,
       volume = {440},
       number = {3},
        pages = {1153-1159},
          doi = {10.1051/0004-6361:20053307},
archivePrefix = {arXiv},
       eprint = {astro-ph/0506458},
 primaryClass = {astro-ph},
       adsurl = {https://ui.adsabs.harvard.edu/abs/2005A&A...440.1153D}
}

@ARTICLE{Opik1924,
       author = {{{\"O}pik}, E.},
        title = "{Statistical Studies of Double Stars: On the Distribution of Relative Luminosities and Distances of Double Stars in the Harvard Revised Photometry North of Declination -31{\textdegree}}",
      journal = {Publications of the Tartu Astrofizica Observatory},
         year = 1924,
        month = jan,
       volume = {25},
        pages = {1},
       adsurl = {https://ui.adsabs.harvard.edu/abs/1924PTarO..25f...1O}
}

@ARTICLE{Arnett:1972,
       author = {{Arnett}, W. David},
        title = "{Advanced Evolution of Massive Stars. II. Carbon Burning}",
      journal = {\apj},
         year = 1972,
        month = sep,
       volume = {176},
        pages = {699},
          doi = {10.1086/151672},
       adsurl = {https://ui.adsabs.harvard.edu/abs/1972ApJ...176..699A}
}

@ARTICLE{Tauris:2013,
       author = {{Tauris}, T.~M. and {Langer}, N. and {Moriya}, T.~J. and {Podsiadlowski}, Ph. and {Yoon}, S.-C. and {Blinnikov}, S.~I.},
        title = "{Ultra-stripped Type Ic Supernovae from Close Binary Evolution}",
      journal = {\apjl},
         year = 2013,
        month = dec,
       volume = {778},
       number = {2},
          eid = {L23},
        pages = {L23},
          doi = {10.1088/2041-8205/778/2/L23},
archivePrefix = {arXiv},
       eprint = {1310.6356},
 primaryClass = {astro-ph.SR},
       adsurl = {https://ui.adsabs.harvard.edu/abs/2013ApJ...778L..23T}
}

@ARTICLE{Willems:2004,
       author = {{Willems}, B. and {Kalogera}, V. and {Henninger}, M.},
        title = "{Pulsar Kicks and Spin Tilts in the Close Double Neutron Stars PSR J0737-3039, PSR B1534+12, and PSR B1913+16}",
      journal = {\apj},
         year = 2004,
        month = nov,
       volume = {616},
       number = {1},
        pages = {414-438},
          doi = {10.1086/424812},
archivePrefix = {arXiv},
       eprint = {astro-ph/0404423},
 primaryClass = {astro-ph},
       adsurl = {https://ui.adsabs.harvard.edu/abs/2004ApJ...616..414W}
}

@ARTICLE{Kramer:1998,
       author = {{Kramer}, Michael},
        title = "{Determination of the Geometry of the PSR B1913+16 System by Geodetic Precession}",
      journal = {\apj},
         year = 1998,
        month = dec,
       volume = {509},
       number = {2},
        pages = {856-860},
          doi = {10.1086/306535},
archivePrefix = {arXiv},
       eprint = {astro-ph/9808127},
 primaryClass = {astro-ph},
       adsurl = {https://ui.adsabs.harvard.edu/abs/1998ApJ...509..856K}
}

@ARTICLE{Dewi:2005,
       author = {{Dewi}, J.~D.~M. and {Podsiadlowski}, Ph. and {Pols}, O.~R.},
        title = "{The spin period-eccentricity relation of double neutron stars: evidence for weak supernova kicks?}",
      journal = {\mnras},
         year = 2005,
        month = oct,
       volume = {363},
       number = {1},
        pages = {L71-L75},
          doi = {10.1111/j.1745-3933.2005.00085.x},
archivePrefix = {arXiv},
       eprint = {astro-ph/0507628},
 primaryClass = {astro-ph},
       adsurl = {https://ui.adsabs.harvard.edu/abs/2005MNRAS.363L..71D}
}

@ARTICLE{HartiganHartigan:1985,
 ISSN = {00905364, 21688966},
 URL = {http://www.jstor.org/stable/2241144},
 author = {J. A. Hartigan and P. M. Hartigan},
 journal = {The Annals of Statistics},
 number = {1},
 pages = {70--84},
 publisher = {Institute of Mathematical Statistics},
 title = {The Dip Test of Unimodality},
 urldate = {2026-10-02},
 volume = {13},
 year = {1985}
}
\bibliographystyle{aasjournal}

% ==========================================================
\appendix
\label{sec:appendix}
% ==========================================================

\setcounter{figure}{0}
\renewcommand{\thefigure}{A\arabic{figure}}

% ==========================================================
\section{Additional population synthesis models}
\label{appMoreModels}
% ==========================================================

\renewcommand{\thefigure}{A1}
\begin{figure}
    \centering
    \includegraphics[width=\linewidth]{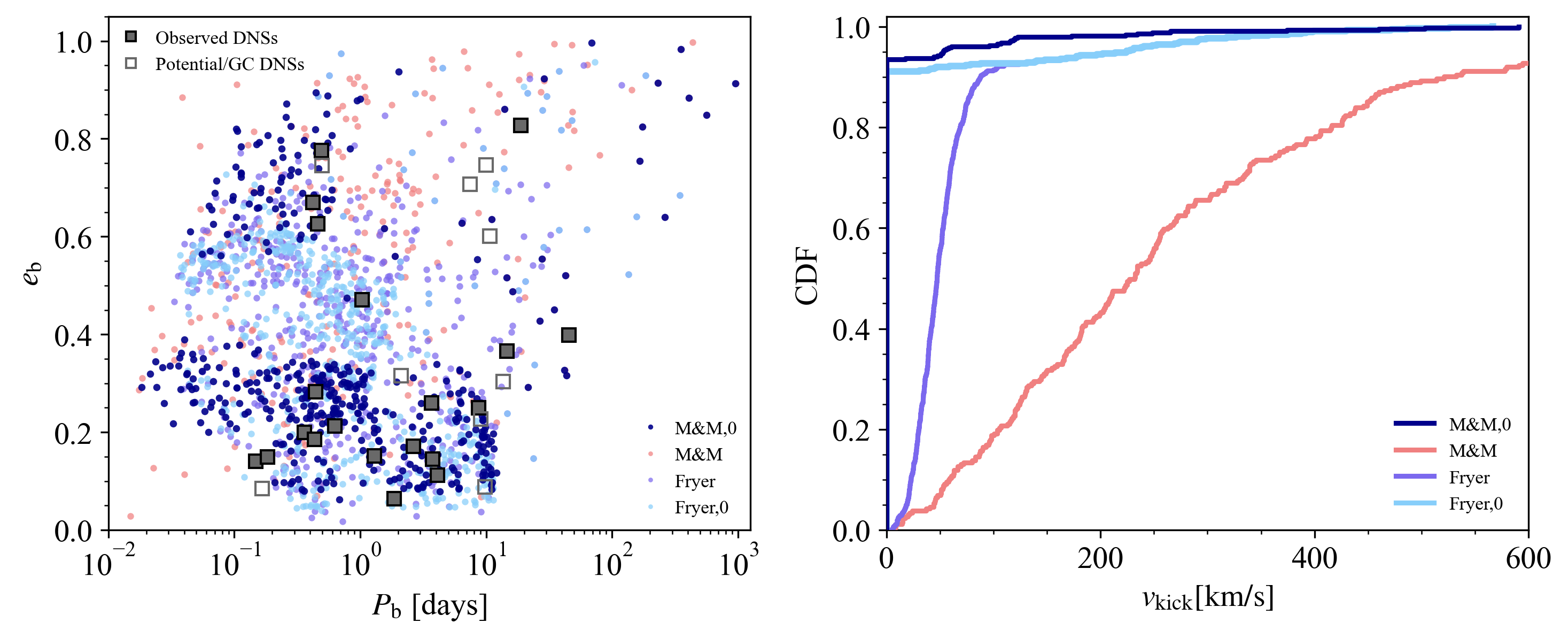}
    \caption{Properties of DNSs for the different remnant mass and kick prescriptions described in Appendix \ref{appMoreModels}: the $\rm M\&M$ model (i, pink dots), $\rm Fryer$ model (ii, violet dots), and $\rm Fryer{,}0$ model (iii, light blue dots). $\rm M\&M{,}0$ (dark blue dots) is presented for comparison. Left panel: period -- eccentricity distribution at DNS birth. The filled gray squares are observations of Galactic DNSs back-integrated to DNS birth, and the empty squares are systems that are associated with GCs or whose nature is debated (see Table~\ref{tab1}). Right panel: cumulative distribution functions (CDFs) of kicks applied to the NS born in the second SN explosion. }
    \label{fig:eP_and_kick_hist_2panel}
\end{figure}

To test how different components of our population synthesis model affect its compatibility with observations, and particularly its ability to reproduce the observed eccentricity bimodality, we perform additional simulations in which we vary key assumptions: 
\begin{enumerate}
  \renewcommand{\labelenumi}{\roman{enumi}.}  
  \item $\rm M\&M$: \cite{MandelMueller:2020} remnant mass prescriptions and kick formalism with the model parameters as in $\rm M\&M{,}0$ and no special treatment of the USSN kick. 
  \item $\rm Fryer$: remnant mass distribution and delayed SN engine prescription from \cite{Fryer:2012}; two-component Maxwellian  kick distribution with $\sigma_{\rm kick,CCSN}=\rm 217 \; km \:s^{-1}$ for core-collapse SNe \citep{DisbergMandel:2025}, and $\sigma_{\rm kick,ECSN}=\rm 30 \; km \:s^{-1}$, $\sigma_{\rm kick,USSN}=\rm 30 \; km \:s^{-1}$ for electron-capture SNe and USSNe (e.g., \citealt{VignaGomez:2018}; see also \citealt{Breivik:2020}; \citealt{Chattaraj:2025} for similar assumptions), respectively.
  \item $\rm Fryer{,}0$: remnant mass distribution and delayed SN engine prescription from \cite{Fryer:2012}; three-component Maxwellian  kick distribution with $\sigma_{\rm kick,CCSN}=\rm 217 \; km \:s^{-1}$ for core-collapse SNe \citep{DisbergMandel:2025}, $\sigma_{\rm kick,ECSN}=\rm 30 \; km \:s^{-1}$ for electron-capture SNe (e.g., \citealt{VignaGomez:2018}), and $\sigma_{\rm kick,USSN}=\rm 0 \; km \:s^{-1}$ for USSNe.
\end{enumerate}

Fig.~\ref{fig:eP_and_kick_hist_2panel} shows the orbital period -- eccentricity distribution (left panel) of models (i)-(iii) immediately after the kick imparted by the second SN explosion (right panel) for binaries that ultimately form DNSs, along with the $\rm M\&M{,}0$ results for comparison. Remnant mass and kick prescriptions without a combination of a mass jump and very low USSN kick fail to reproduce the possible eccentricity bimodality suggested by the back-integrated Galactic DNS sample.

\renewcommand{\thefigure}{A2}
\begin{figure}
    \centering
    \includegraphics[width=\linewidth]{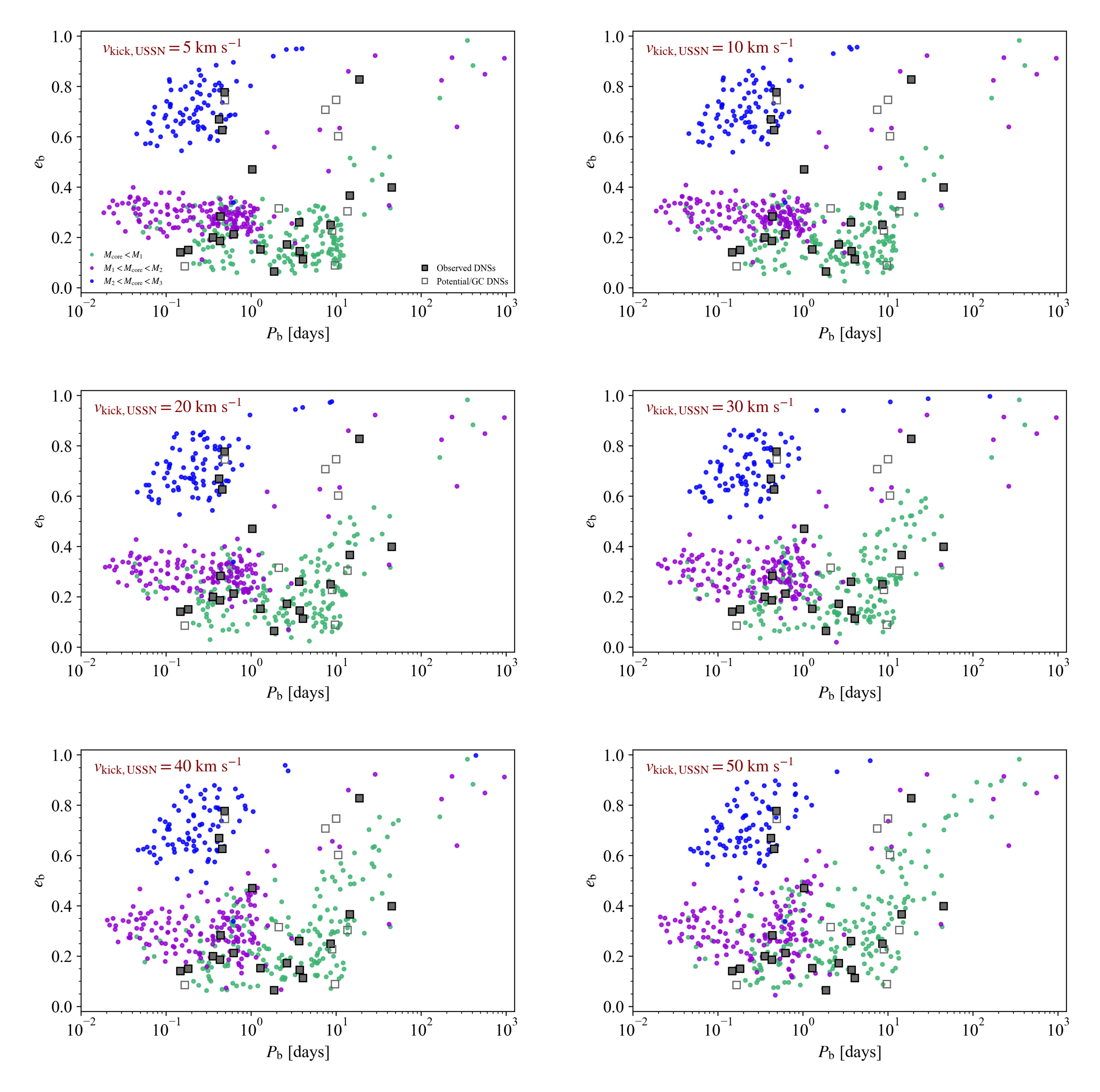}
    \caption{Orbital period -- eccentricity distributions for population synthesis simulations run with the \cite{MandelMueller:2020} recipes assuming USSN kicks of magnitude $v_{\rm kick,USSN}$, labeled in the top left of each panel. The notation for simulated and observed data is the same as in Fig.~\ref{fig:two_panel_e_prob_masses}.}
    \label{fig:Different_USSN_kicks}
\end{figure}

In addition to these three different prescriptions, we run six simulations with similar model parameters to the $\rm M\&M{,}0$ model, varying the kick imparted by USSNe from $v_{\rm kick,USSN} = \rm  5 \:km\:s^{-1}$ to $v_{\rm kick,USSN} = \rm  50 \:km\:s^{-1}$. In Fig.~\ref{fig:Different_USSN_kicks}, we present orbital period–eccentricity distributions, analogous to the top left panel of Fig.~\ref{fig:two_panel_e_prob_masses}, for these simulations. When the second NS receives a non-zero natal kick at birth, the post-SN eccentricity is no longer set solely by the \cite{Blaauw:1961} kick; instead, the additional natal kick broadens the eccentricity distribution. As the USSN kick magnitude increases, the eccentricity gap becomes progressively populated.

% ==========================================================
\section{Statistical significance of the gap}
\label{appGapSignificance}
% ==========================================================

The back-integrated Galactic DNS observations suggest a scarcity of systems at intermediate eccentricities, with only one system with eccentricity of $\sim 0.5$ after back integration. We evaluate whether this apparent deficit provides statistically significant evidence for multimodality given the small sample size.

We first use the dip test of \cite{HartiganHartigan:1985}, which quantifies how strongly the empirical cumulative distribution departs from the closest unimodal cumulative distribution, to test the eccentricity distribution for multimodality. We apply it to the back-integrated eccentricities of the $N=20$ confident non-GC DNSs and find the size of this departure to be $D\simeq0.042$, and the corresponding probability of obtaining a departure at least as large as the observed value under the null hypothesis of unimodality to be $p\simeq0.99$, indicating that the current sample does not provide statistically significant evidence for multimodality.

We then adopt a smooth eccentricity distribution that contains no gaps and approximately matches the observed CDF at $e \gtrsim e_{\rm min} =0.05$. This choice is intended as an illustrative, smooth benchmark rather than a unique physically-motivated prescription; however, a nonzero lower eccentricity bound can be qualitatively motivated by a minimum eccentricity imparted by neutrino mass loss in the second SN. Specifically, we consider a CDF of the form 
\begin{equation}
\begin{split}
P(E \le e) = 1 -\frac{\ln(e)}{\ln(e_{\rm min})},
\label{eq:CDF}
\end{split}
\end{equation}
as shown in Fig.~\ref{fig:CDF_plot}, and compare it to the observational CDF, and to CDFs computed with the population synthesis data of the $\rm M\&M{,}0$ model and the rest of the models described in Appendix~\ref{appMoreModels}.
\renewcommand{\thefigure}{B}
\begin{figure}
    \centering
    \includegraphics[width=\linewidth]{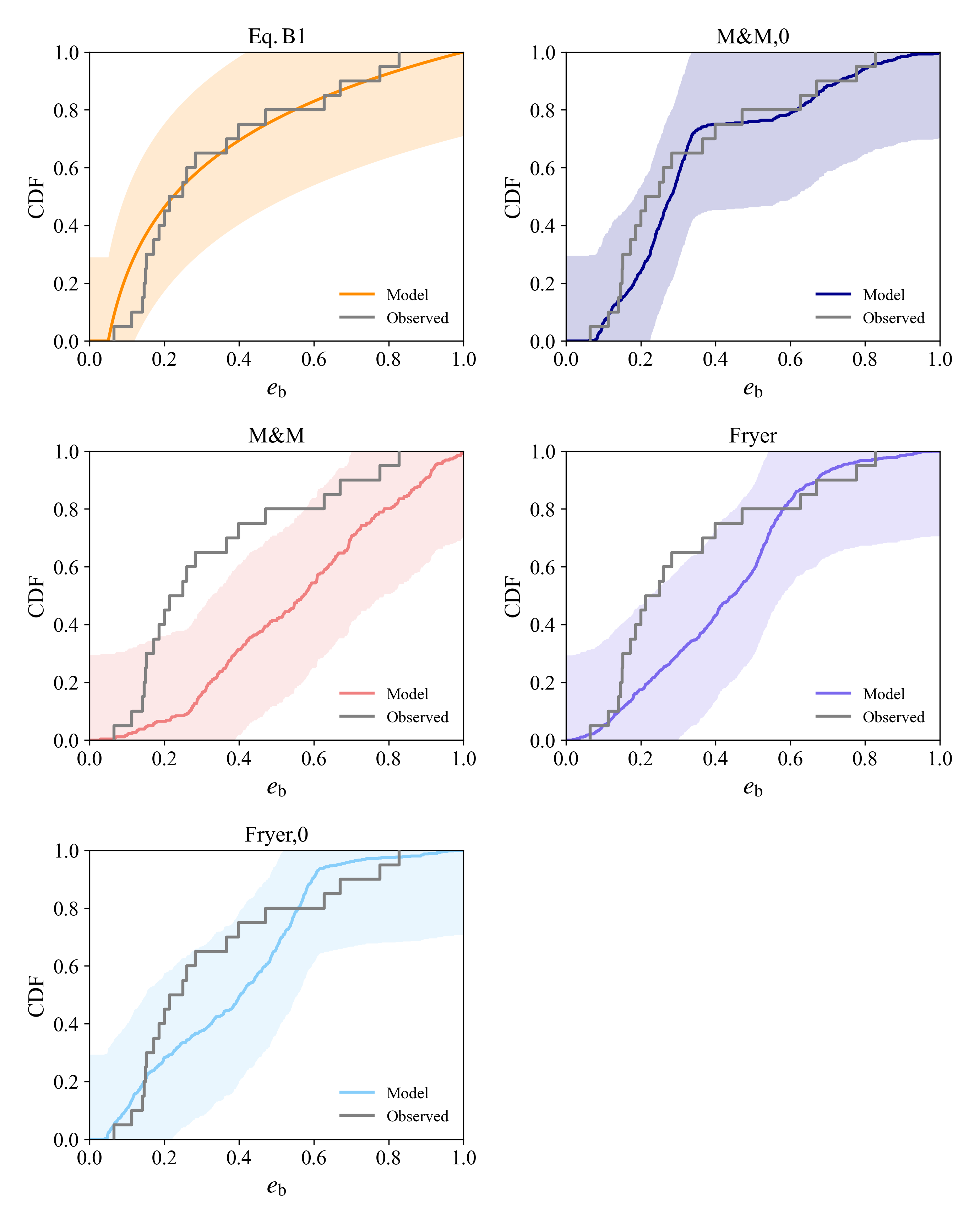}
    \caption{CDFs of Galactic DNS orbital eccentricities compared to finite-sample model predictions. Each panel shows one model: the gray step curve is the observed empirical CDF of the 20 confident non-GC Galactic DNSs, and the colored curve is the model CDF. The orange curve represents the CDFs of a potential continuous distribution (equation \ref{eq:CDF}), while the dark blue, pink, violet and light blue curves are the CDFs of the $\rm M\&M{,}0$, $\rm M\&M$, $\rm Fryer$, and $\rm Fryer{,}0$ models, respectively. The shaded region is the $95\%$ uncertainty interval on model predictions for samples of size $N = 20$. The observed eccentricity distribution strongly disfavors $\rm M\&M$ and disfavors $\rm Fryer$ at the $5\%$ level.}
    \label{fig:CDF_plot}
\end{figure}

First, we compare the empirical eccentricity CDF of the observed, back-integrated DNSs to the smooth eccentricity distribution using a Kolmogorov–Smirnov (KS) test \citep{massey1951kolmogorov}. We find that the observed CDF is consistent with the smooth distribution, with a KS $p-$value of $\simeq 0.12$, implying that if the continuous eccentricity distribution is correct, a discrepancy at least as large as observed would occur in $\sim 12 \%$ of samples of the same size due to statistical fluctuations. Thus, the data does not provide strong evidence to reject a smooth, gap-free eccentricity distribution at conventional significance thresholds. The $95\%$ uncertainty interval, though, disfavors $\rm M\&M$ and $\rm Fryer$, while favoring the no-kick models (Fig.~\ref{fig:CDF_plot}). 

We next compute the probability that, under the same smooth distribution, a sample of size $N$ contains an empty interval at least as large as the scarcely populated intermediate eccentricity regime anywhere in the distribution. The probability of observing a system in this range is  
\begin{equation}
\begin{split}
p_{\rm gap} = P(e_{\rm 1} \le E \le e_{\rm 2}) = \frac{\ln(e_{\rm 1}/e_{\rm 2})}{\ln(e_{\rm min})},
\label{eq:p_gap}
\end{split}
\end{equation}
which results in $p_{\rm gap} \simeq 0.124$ for the intermediate-eccentricity interval $[0.4,0.58]$. In other words, the observed empty interval contains $12.4\%$ of the probability mass of the smooth distribution. To account for the fact that the boundaries were identified after inspecting the data, we consider the largest internal spacing in a sample rather than the probability of finding no systems in this particular interval. Transforming the eccentricities according to the CDF in Equation~\ref{eq:CDF} makes them uniformly distributed on $[0,1]$ under the smooth-distribution hypothesis. For $N$ uniformly distributed points, we calculate the probability that at least one of the $N-1$ internal spacings is as large as, or larger than, the currently observed gap in measured eccentricities.
\begin{equation}
\begin{split}
p_{\rm max,gap} = 1 - \sum _{j=0}^{\min\left(N-1,\left\lfloor 1/p_{\rm gap}\right\rfloor\right)} (-1)^{j} \binom{N-1}{j} (1-jp_{\rm gap})^N.
\label{eq:p_max_gap}
\end{split}
\end{equation}
For the $N=20$ confident observed non-GC DNSs we obtain $p_{\rm max,gap}\simeq 0.86$, i.e., approximately $86\%$ of samples of 20 systems contain an empty interval with at least as much probability mass as the sparsely populated intermediate eccentricity region somewhere in the distribution. This implies that the absence of systems in this region is not statistically unusual under a smooth distribution given the current sample size. 

Therefore, our calculations indicate that the current sample does not provide statistically significant evidence for bimodality, and that the observed eccentricity distribution is more consistent with models in which the second-born NS receives a small or negligible natal kick. Only a much larger observed sample, potentially enabled by SKA \citep{Braun:2019}, which is expected to detect dozens to hundreds of new Galactic DNSs, will reveal whether an eccentricity bimodality really exists and constrain the typical natal kicks imparted in the second SN.

\end{document}